\documentclass[10pt,journal,compsoc]{IEEEtran}

\usepackage{blindtext}
\usepackage{graphicx}
\usepackage{eqnarray,amsmath}
\usepackage{multirow}
\usepackage{amssymb}
\usepackage[table]{xcolor} 
\usepackage{footmisc}
\usepackage{multirow}
\usepackage{makecell}
\usepackage{subcaption}
\usepackage{lipsum}

\usepackage{float}
\usepackage{color}
\usepackage{comment}
\usepackage{amsmath}
\usepackage{amsfonts}
\usepackage{esvect}
\usepackage{url}
\usepackage{hyperref}
\hypersetup{
	colorlinks=false,
	pdfborder={0 0 0},
}

\usepackage[ruled,vlined,linesnumbered]{algorithm2e}
\usepackage{diagbox}
\usepackage{slashbox}
\usepackage{ragged2e}

\begin{document}

\title{Block Chain and Internet of Nano-Things for  Optimizing Chemical Sensing in Smart Farming}

\author{Dixon Vimalajeewa, Subhasis Thakur, John Breslin, Donagh P. Berry, Sasitharan Balasubramaniam 
	\IEEEcompsocitemizethanks{
		\IEEEcompsocthanksitem D. Vimalajeewa and S. Balasubramaniam are  with the Telecommunications Software and Systems Group, Waterford Institute of Technology, Waterford, Ireland, E-mail: dvimalajeewa@tssg.org and sasib@tssg.org.
		
		\IEEEcompsocthanksitem S. Thakur and J. Breslin are with the National University Galway, Ireland, E-mail : subhasis.thakur@insight-centre.org and john.breslin@nuigalway.ie.
		\IEEEcompsocthanksitem D. P. Berry is with the Teagasc, Animal \& Grassland Research and Innovation Centre, Moorepark, Cork, Ireland, E-mail : Donagh.Berry@teagasc.ie.
		
		This work was supported by the Science Foundation Ireland (SFI) projects PrecisionDairy (ID: 13/1A/1977) and VistaMilk (ID: 16/RC/3835), and the Horizon 2020 GenTORE project.
			
		E-mail: dvimalajeewa@tssg.org
	}
	\\

}


\IEEEtitleabstractindextext{%

\begin{abstract}
\justify{The use of Internet of Things (IoT) with the Internet of Nano Things (IoNT) can further expand decision making systems (DMS) to improve reliability as it provides a new spectrum of more granular level data to make decisions. However, growing concerns such as data security, transparency and processing capability challenge their use in real-world applications. DMS integrated with Block Chain (BC) technology can contribute immensely to overcome such challenges. The use of IoNT and IoT along with BC for making DMS has not yet been investigated. This study proposes a BC-powered IoNT (BC-IoNT) system for sensing chemicals level in the context of farm management. This is a critical application for smart farming, which aims to improve sustainable farm practices through controlled delivery of chemicals. BC-IoNT system includes a novel machine learning model formed by using the Langmuir molecular binding model and the Bayesian theory, and is used as a smart contract for sensing the level of the chemicals. A credit model is used to quantify the traceability and credibility of farms to determine if they are compliant with the chemical standards. The accuracy of detecting the chemicals of the distributed BC-IoNT approach was $\geq90\%$ and the centralized approach was $\leq80\%$. Also, the efficiency of sensing the level of chemicals depends on the sampling frequency and variability in chemical level among farms.
}
\end{abstract}

}

\maketitle

\IEEEdisplaynontitleabstractindextext
\IEEEpeerreviewmaketitle

\IEEEraisesectionheading{\section{Introduction}\label{intro}}
Smart farming (SF) is transforming farming practices by integrating modern Information and Communication Technologies (ICT) to achieve greater productivity through sustainable practices aiming to address the world food demand for increasing world population that is expected 9.7 billion by 2050. In doing so, minimizing SF carbon footprint (i.e., total greenhouse gas (GHG) emission), today, takes the highest priority. That is because, agriculture is one of the major contributors to global GHG emission. For instance, the GHG emission from the agri-sector is 12.56\% of the global GHG emission in 2016 and expected to be around 58\% by 2050 \cite{37}. It is, however, greatly emphasized that GHG emission can be cut by 40\% through incorporating modern technological advances such as nanotechnology with SF practices \cite{33, 35, 36}. Therefore, the SF sector is now widely adopting various farming practices such as control delivery of chemical fertilizers, herbicides and pesticides, incorporating with modern technologies for reducing GHG emissions, thereby improving the sustainability of the SF practices. 

With the progressive adoption of modern ICT in SF, agri-tech has become further advanced, resulting in novel and intelligent services. The recent advances in the Internet of Things (IoT), coupled with modern ICT, facilitates continuous monitoring of spatial and temporal variability that exists from sensors integrated into various SF application domains such as food production and supply chain. In particular, sensor technologies have undergone dramatic advancements due mainly to the developments in other supporting fields such as nanotechnology, which enables monitoring of molecules at a fine granular scale. The emerging field of nano-communications allows communication and networking between devices to be developed from nano-scale components \cite{17}. The integration of nano-communications with IoT has led to a new paradigm known as the Internet of Nano-Things (IoNT), empowering the potential of creating a broader spectrum of data that can take the SF applications described above to unprecedented levels. Today, IoT and IoNT coupled with Machine Learning (ML) and Artificial Intelligence (AI)-based decision-making systems (DMSs) promise to create high value and innovative services to generate timely and accurate insights required for operating more environmental friendly SF practices \cite{39, 41}. 

Effective data processing for the generation of timely accurate insights required for operating more sustainable SF practices is, however, becoming highly challenging through cloud-based DMSs only. That is because the greater adoption of IoT and IoNT devices enables collecting massive datasets though, it leads to exceeding resource capacities, limited scalability, and extended latency. Therefore, cloud-based DMSs were extended to Fog and then Edge computing-based DMSs. Cloud-, Fog-, and Edge-computing paradigms based DMSs provide information to make timely decisions for performing farm operations effectively.  Nonetheless, several concerns have been raised with regards to such DMSs in that they may not be providing adequate support for addressing certain criteria effectively \cite{11, 38}. Most specifically, traceability, transparency and trustworthiness of the information provided for making accurate decisions to operate effective farming practices. That is because, with the rapid advancement in SF, the main aim of SF is now, not only to intensify sustainable farm production but also manage an effective food supply chain. The high demand for food has created greater market competition, and consequently has increased the availability of similar products for different brands.  Hence, consumers are now more conscious about being more aware of production quality and eco-friendliness for the products they purchase. Information for ensuring, for instance, quality and safety standards is, however, often lacking in DMSs which are currently in use. This is mostly due to the data privacy and issues and fraud behaviors, where producers are reluctant to share data such as the use of synthetic chemicals in the food production process for ensuring food safety and quality and also, lack of communicability between food production and supply chain practices. Consequently, these factors bring several disadvantages such as increasing food and farm input wastage, lowing food quality and safety, challenging human health and eco-friendliness and eventually leading to a higher carbon footprint value, and in particular, the greater potential of SF data is significantly underutilized in optimizing SF practices. Therefore, considerable attention has been expended today on the urgency of restructuring the existing DMSs (i.e., Cloud-, Fog-, and Edge-computing paradigms based DMSs) into fully distributed DMSs that enable providing more transparent, secure, and traceable insights to make more valid, trustworthy and real-time decisions. 

Block Chain (BC) technology has become one of the promising solutions to build fully distributed DMSs equipped with such features~\cite{18}. BC technology enables operating an immutable and distributed ledger, so that information can be made securely accessible to all participants (e.x., consumers and food producers and distributors in the food supply chain) in the DMSs, enabling them to communicate directly without depending on intermediates which leads to security issues~\cite{28}. Consequently, DMSs incorporated with BCs enable producing transparent, traceable, and trustworthy decisions. The BC technology has been widely used in the financial sector, though its significance has not yet been fully realized in many other fields where numerous benefits may exist. Therefore, this gap is {\bf the key motivation} for conducting this study, focusing on how the smart farming sector can be powered by BC technology to help to achieve sustainable farming practices and effective food supply chain. Although the incorporation of existing DMSs with BC technology has a greater potential in developing such fully distributed DMSs, development of such DMSs has not much been taken into consideration in the SF sector \cite{38}.

{\bf The aim of this study} is, therefore, to explore how IoNT, IoT and BC can be used together to develop a DMS based on an application that detects the level of chemical usage in farmlands. Then making chemical usage information available to stakeholders in the food supply chain in a more secure, transparent, and traceable manner to be able to make more trustworthy decisions. The use of different chemicals such as synthetic fertilizers and both herbicides and pesticides in SF is a common practice, for instance, to maintain optimal soil quality, increase the yield and quality of crop and vegetables and prevalence of diseases. Today, a number of SF practices do not achieve these goals, mainly due to the use of chemicals with insufficient monitoring, data processing and decision-making facilities. Consequently, these practices bring critical challenges such as reduced productivity and production quality and in particular, increase in GHG emission. Moreover, market demand for low-quality food decreases as consumers are now more health-conscious, as such they mostly prefer buying only high quality and safety of products (e.x., organic food). Altogether, these factors lead to huge food wastage and economic loss and eventually adverse environmental impact (i.e., higher carbon footprint) in the modern SF and food supply chain. Therefore, having a mechanism not only to sensing the optimal chemical requirements for sustainable intensification of farm production with minimal carbon footprint, but also making such practices transparent to the stakeholders in the food supply chain in a real-time manner is essential to overcome such challenges effectively. Consequently, this will enhance the usability of SF data for optimizing farming practices. Moreover, producers will be able to use the decisions (or insights) generated by such systems to place timely management practices for optimal chemical usage, minimizing the carbon footprint while consumers can be more aware of the quality and safety of the food they consume. Therefore, {\bf the main contribution of this study} is developing a BC-based approach summarized below for identifying the chemical levels (or usage) in farmlands and then explore its usability as a platform for strengthening trust between producers and customers through making chemical usage data available to customers as a color token. 

\begin{itemize}
    \item[a.] {\bf Nanotechnology in SF monitoring:} Explore the functionality and usability of affinity-based nano-sensors for sensing the availability of certain chemicals in soil.   
    
    \item[b.] {\bf Intelligent data processing:} Incorporate data collected from nano-sensors with IoT devices and then using the Bayesian probability updating approach for deriving insights regarding the chemical levels in the farm or crop.
    
    \item[c.] {\bf Fully distributed DMS:} Insights from step [b] are combined with the BC technology to demonstrate the effectiveness of the BC-enabled approach for detecting the use of chemicals.
    
    \item[d.] {\bf Decision-making:} Introduce a token-based credit system to ensure the credibility and traceability of farms being compliant with chemical standards in their production processes. 
    
    \item [e.]  {\bf Assessment of trustworthiness of decisions:} Discuss the assessment of credibility and traceability of the farming processes, including benefits and challenges of the proposed BC-IoNT system.
\end{itemize}

The remainder of the paper is organized as follows. Section~\ref{sec2} discusses related works and Section~\ref{sec3} explains the system model. In Section~\ref{sec4}, the system model described in Section~\ref{sec3} is applied on simulated data to detect the level of chemicals. Section~\ref{sec5} discusses the way of ensuring traceability and credibility of the proposed BC system, and Section~\ref{sec6} concludes the paper.

\section{Related Work}\label{sec2}
We herein discuss the use of IoNT, IoT and BC in developing different DMS and challenges, and then the significance of the present study with respect to the already existing solutions overcoming those challenges. 

\subsection{ IoNT and IoT based DMS}
In  IoNT and IoT based DMS, data sensed by nano-sensors are aggregated at nano-routers to send to macro-scale devices (e.x., IoT). That is because these IoNT devices can, however, perform limited tasks due to limited resources, so that they are usually operated in integration with IoT-devices. Therefore, IoNT integrated with IoT improve the potential of enhancing, and also scaling up, the services of existing systems such as controlled delivery of drugs \cite{17}. 

Currently, such integrated networks based DMSs mostly operate under Cloud-computing paradigm \cite{11}. Extended latency, network congestion, and safety of data are some of the critical problems that limit the use of this computing method in many applications. Alternatively, distributed computing approaches based DMSs like Fog- and Edge-enabled systems coupled with state-of-the-art ML techniques such as federated learning and deep learning have been proposed. Such systems can effectively overcome those issues while ensuring data privacy and security, thus enhancing the reliability and timeliness of outcomes \cite{3, 38}. 
 
However, trustability, integrity, and functional incompatibility are some of the most important factors which hamper the interoperability of such advanced computing systems, technologies, and devices. Consequently, those computing systems operate in isolation and are poorly scalable, so that their full potential, as well as the collected data, are significantly under-utilized. Therefore, the urgency of developing alternatives to handle these issues has gained considerable attention from the wider research community. A number of recent studies such as~\cite{14, 27, 28} have emphasized that BC and IoT are growing together and codependent, so that they have a greater potential of overcoming such issues. 

\subsection{Block Chain(BC) }\label{bc}
The distributed ledger technology (DLT) is a fully decentralized peer-to-peer (P2P) method used for recording transactions (i.e., data) in an immutable ledger, with the mechanism for processing, validating, authorizing transactions (or decisions made by DMS in data prcessing). Block chain (BC) is an application of DLT for securely storing data and also known as \textit{'internet of value'} \cite{28}. BC generally consists of three components; blocks containing transaction data, a P2P network for direct communication, and a shared ledger for distributed data storage. Each block contains data with a hash value (a unique identification key) and a pointer to the hash of the previous block. A consensus algorithm is used for creating a new block and then appending it to the ledger. The Proof-of-Work (PoW) mechanism is the most commonly used consensus algorithm \cite{18,26}. In PoW, a hash key is allocated to a block through a mathematical puzzle, which is hard to solve, but easy to verify. Deriving a unique hash key is a computationally heavy task so that nodes which have sufficient resources are used for that and known as mining nodes. 

Today, different versions as well as types of BCs are available for various applications. The study \cite{26} stated two versions of BC as BC.01 and BC.02, which uses cryptocurrency and smart contracts (explain later), respectively. There are mainly three types of BC termed consortium-, private-, and public-BC \cite{28}. While the consortium-BC is a formed combining a set of BCs \cite{8}, a trusted central entity controls the private BC and termed permissioned BC \cite{6}. The public-BC is called permissionless BC because there is nobody to control it and also anybody can join (or leave) the BC network at any time. Selection of the most suitable BC depends on the application requirements though the study \cite{26} emphasized that there are unique properties such as irreversibility, traceability, anonymity, security and transparency that enable the use of BC in a vast range of applications. Some of their most prominent applications are intelligent management \cite{25}, smart transportation \cite{24}, agriculture \cite{14}, and healthcare \cite{7}.

\subsection{Integration of IoT with BC}
Many attempts have been made to incorporate BC technology with IoT-based systems. The intention of forming such integrated systems can be mainly seen under three categories: overcoming the issues in existing IoT-based DMS platforms, optimizing the resources for efficient BC operations, and testing usability and improving the reliability of DMSs in different applications. 

Considering the studies conducted to overcome the issues of the existing IoT-based platforms, the study\cite{3}, for instance,  proposed a BC-enabled federated learning ({\it BlockFL}) method to overcome general issues in federated learning-based distributed ML systems such as single point failure. This study also investigated end-to-end learning competition aiming to find optimal block generation rate. Meanwhile, an analytical model proposed in \cite{1} discussed the optimal deployment of full functional BC nodes for a BC-enabled wireless IoT system, minimizing data security issues. Finding sufficient resources for employing BC-powered IoT systems is a critical challenge that has been taken into a broader consideration. Data management and access control methods for BC-enable IoT systems given in \cite{4} explained how time-series data could be stored at the edge of the IoT network for effective processing. The study \cite{5}, for instance, proposed an auction-based resource allocation method in connection with Cloud/Fog computing while a BC-based big data-sharing platform for resource-limited edges was developed in \cite{10} by considering the challenges in deploying BC in edge devices. With regards to enhancing the performance of BC systems, BC-enabled edge computing approach was proposed in \cite{6}, aiming to ensure data privacy and energy security for power smart grid network. The work presented in \cite{8} also explains how to manage power for plug-in electric vehicles in smart grids. Moreover, consortium BCs have been widely used for enhancing BC performance. The study \cite{7} discussed a consortium BC-based mechanism for improving the accuracy and effectiveness of disease diagnosis in health-care. 

Some studies, however, warned that these integrated systems could have unfavorable responses. The study \cite{28}, for instance, warned that this integration could also create unnecessary computational overhead and may not generate any tangible benefits. Therefore, \cite{25} and \cite{28} recommended to conduct an initial case study to make sure that integration with BC is necessary, proposing a checklist to conduct such a feasibility study.

The evidence already provided, however, emphasizes the significance of BC-powered IoT systems, highlighting the performance of the existing IoT systems which can be empowered with improved scalability. On the other hand, it is already proven that IoNT can contribute to improving the performance of the IoT-based systems \cite{32}. For instance, \cite{35} stated that 30\% of Irish GHG emission is from agriculture and of which CH$_4$ and N$_2$O accounts nearly 2/3 and 1/3, respectively. In addition, \cite{35} also suggested that adopting various farming practices such as extended grazing, energy efficiency and Nitrogen efficiency to reduce GHG emission. Besides, the study \cite{33} emphasized that the significance of using IoNT in agriculture, highlighting that the IoNT-based SF can significantly contribute to reducing the GHG emission. The study \cite{32} anticipated that 2Gt CO$_2$ emission could be reduced per year through implementing IoNT-based SF practices. On the other hand, according to \cite{34}, attempts such as intelligent monitoring, timely decision making and need-based farm input management strategies have greater potential to save 1 billion 1MW of energy, and on average, increase 897 kg/Ha of land yield. Consequently, this would contribute to saving up to \$110 billion by 2030. However, there is no evidence that any attempt has been made so far to integrate IoNT-based approaches into BC-powered IoT systems. Therefore, this study aims to fill that gap by exploring how IoNT and IoT network can be powered by BC mechanism. The next section presents a system model to explain how IoNT can be used in BC-powered IoT system.

\begin{figure*}[!t]
\centering
  \includegraphics[width= .9\linewidth, height = 7.5cm]{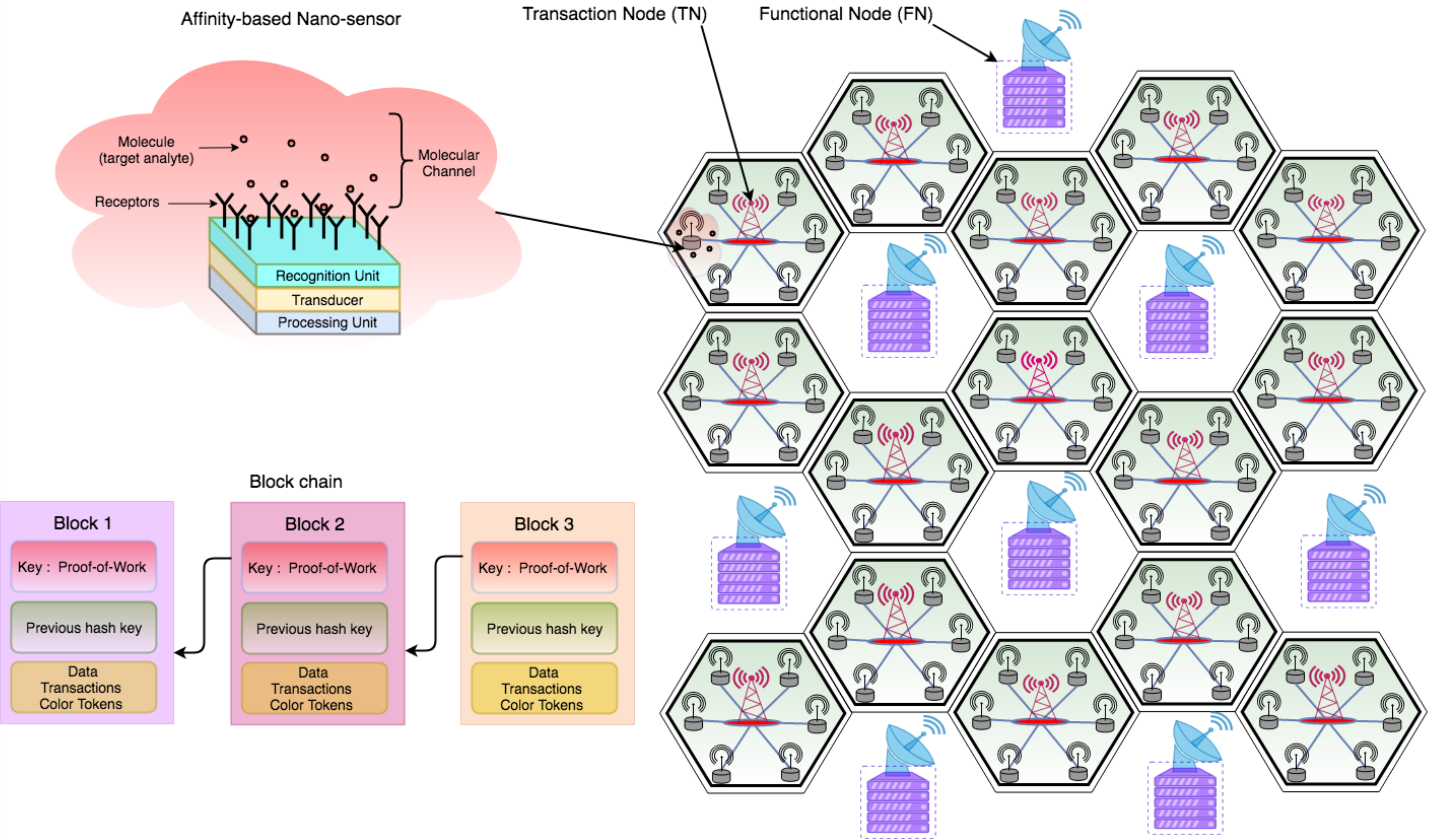}
  \vspace{-1.0em}
  \caption{ Overall architecture of the system model and a sample BC for five transactions.}
    \vspace{-0.5em}
\label{fig-01}
\end{figure*}
\section{System Model}\label{sec3}
The system model is an integration of the functionalities of four components which includes: affinity-based IoNT sensors, IoT devices (gateways/transaction nodes (TN)), miners (functional nodes (FN)), and the BC network. Figure \ref{fig-01} illustrates the system model architecture, including IoNT sensor and an example BC network. The IoNT sensors are connected to each gateway, and it is assumed that IoT devices are connected with each other as well as with the FNs. The functionality of each component in the system is explained in the following subsections. 
\subsection{ Affinity-based IoNT sensors}\label{IoNT}
Among the different types of IoNT sensors, affinity-based nano-sensors are the most commonly used type for monitoring the presence of a specific chemical in a medium. The selective binding between molecules in a chemical sample and target analytes is the primary mechanism used for detecting the presence of that chemical~\cite{17}. The role of the affinity-based nano-sensors is, therefore, to facilitate binding target analytes with bio-molecules (known as receptors), which are functionalized on the nano-sensor surface. As a result, the sensor generates a signal reflecting the abundance of target analytes as a variation in the voltage. Based on the affinity sensor illustrated in Figure \ref{fig-01}, when target analytes bind with the receptors, the following sequence of operation occurs:

\begin{itemize}
    \item [1.] The recognition unit selectively detects the target analytes.
    \item [2.] The transducer converts the recognized events to processable signals in the form of electrical pulses.
    \item [3.] The processing unit extracts the insights encoded in the signal.
\end{itemize}
The next section explains the procedure for deriving insights, which is the analyte concentration from the sensor signal.  

\subsubsection{Deriving analyte concentration through affinity-based nano-sensors signal}\label{affsnsr}
The Langmuir model \cite{15}, formulates a simple $1:1$ interaction between two molecules, is used to describe the functionality of the affinity sensor. The model assumes that all binding sites are equivalent and independent. When the binding interaction occurs between an analyte $A$ (i.e., target molecule) and a receptor molecule $B$, it forms a chemical complex $C$ (i.e.,  $A + B \rightleftharpoons C$). Since the number of receptors ($[B]$) is fixed on the sensor surface, the concentration of $A$ is proportional to $[C]$. This means that a change in the sensor signal (say $R$) is proportional to the concentration of $C$ and, therefore, the concentration of $A$. In the present study, $R$ is computed in response units ($RU$), assuming $[B] = B_{max}$ is fixed \cite{22}. The procedure for deriving the formula for extracting information about the analyte concentration (i.e. $[A]$) is explained below in four steps.

\begin{itemize}
    \item [1.] Considering the first order kinetics of the chemical reaction between $A$ and $B$, the rate of change in $[C]$ and $[A]$ can be written as follows;
        \begin{eqnarray}\label{af_kinetics}
            \frac{d\left[C\right] }{dt} &=& \displaystyle k_a  \left[A \right] \left[B\right] - k_d \left[C\right]  \quad \text{association} \label{af_kinetics1},\\ 
            \frac{d\left[A\right]}{dt} &=& \displaystyle -k_a \left[A\right] -k_d \left[C\right]  \quad \text{disassociation} \label{af_kinetics2}, 
        \end{eqnarray}
        where $k_a$ and $k_d$ are the association and disassociation rates, respectively.\\
        As $R \propto \left[C\right]$, the maximum $R$, $R_{max} \propto B_{max}$ at any time $t$, will result in free receptor concentration $\left[B\right] = B_{max} - \left[C\right]$, i.e., $ R \propto R_{max} - R_a$. Therefore, the equations can be re-written as: 
        \begin{eqnarray}\label{af_response}
            \frac{dR_a}{dt} &=&\displaystyle  k_a  \left[A \right]  (R_{max} - R_a) - k_d R_a \label{af_response1}, \\
            \frac{dR_d}{dt} &=& \displaystyle \left[A\right] -k_d R_d.\label{af_response2}
        \end{eqnarray}
        
        \item [2.] The expressions for sensor response $R$ in the association and disassociation phases ( $R_a$ and $R_d$) are derived by analytically solving the two differential equations \ref{af_response1} and \ref{af_response2} ( note: $\left[A \right] = 0$ in the disassociation phase ). 
        \begin{eqnarray} \label{af_eq}
            R_a &=& \displaystyle \frac{k_a \left[A\right] R_{max}}{(k_a \left[A\right] + k_d)}\left(1 - e^{-t (k_a \left[A\right] + k_d)}\right), \\
            R_d &=& \displaystyle R_{d0} e^{-k_d t},
        \end{eqnarray}
        where $R_{d0}$ is the level of signal at the end of the association.
    
        \item [3.] When $A + B \rightleftharpoons C$ reaches its equilibrium state, $R_a$ is at its maximum (say $R_{eq}$). Therefore, $R_{eq}$ is derived from \ref{af_eq} as:
        \begin{equation}\label{r_eq}
            R_{eq} = \displaystyle R_{max} \frac{ \left[A\right] }{\left[A\right] + k_D }, 
        \end{equation}
        where $ k_D = \frac{k_d}{k_a}$ is the affinity constant which represents the strength of the affinity between $A$ and $B$ ($k_D << 1$ means the higher affinity ).
        \item [4.] The response factor ($RF$), which represents $R_{eq}$ relative to $R_{max}$, is computed as:
        \begin{equation}\label{rf}
            RF = \frac{R_{eq}}{R_{max}} = \frac{ \left[A\right] }{ \left[A\right] + k_D },
        \end{equation}
        when $k_D = \left[A\right]$, $RF = 0.5 $. This means $50\%$ of receptors (i.e.,$B$ molecules) are occupied by $A$ molecules at the equilibrium stage. This reaches $100\%$ when the sensor becomes saturated. 
\end{itemize}
To fit a response curve (i.e., $RF$ model), the response factor formula (\ref{rf}) is used for computing $RF$ values for a range of chemical concentration $[A]$ values. The $RF$ model represents the relationship between $RF$ and $[A]$ and enables identification of the sensor response region where a significant change in $RF$ can be obtained in response to the change in $[A]$. Also, this model can be used to derive the $[A]$ corresponding to any $RF$ value. Finally, this $RF$ value is sent to a gateway node to which the nano-sensor is connected to. Algorithm \ref{Al1} presents the process of computing $RF$ for a given $[A]$. A detailed discussion about deriving the $RF$ model can be found in \cite{15} and \cite{16}.

\begin{algorithm}[!t]
\DontPrintSemicolon
\SetAlgoLined
\SetKwInOut{Input}{Input}\SetKwInOut{Output}{Output}
\Input{$k_a, k_d, R_{max}$, \&  $[A]$}
\Output{$RF$}
 Initialization for the association phase\ ;
 $R_{(a,t = 0)} = \epsilon, t= 0,  \quad \text{and} \quad  \varepsilon_R = 1 \times 10^{-5}$ \;
\BlankLine
\While{$\epsilon \geq \varepsilon_R $}{
    $R_{a,t} = R_a(t, k_a, k_d, [A])$ \;
    $\epsilon = |R_{a, t} - R_{a, t-1}|$ \;
    $t = t + 1$\;
}
\BlankLine
 Initialization for the disassociation phase\ ; $R_{d,t} = R_{a,t}, t = t, $ \;
\While{$\epsilon \geq \varepsilon_R $}{
    $R_{d,t} = R_d(t, k_d)$ \;
    $\epsilon = |R_{d, t} - R_{d, t-1}|$ \;
    $t = t + 1$\;
}
$RF = \text{Max}\{R_{a,t}, R_{d,t}\}/R_{max}$\;
\caption{Nano-sensor Response Factor ($RF$) }
\label{Al1}
\end{algorithm}

\begin{table}[!t]
  \begin{center}
    \caption{Response Classes (RC)}
    \begin{tabular}{cccccc} 
    \hline
      \textbf{RF ($\%$)}& $<$20 &  20-40 & 40-60 & 60-80 & $>$80\\
      \hline
      \textbf{RC} & A & B & C & D & E  \\
      \hline
    \end{tabular}
  \label{tab1}
  \end{center}
\end{table}

\subsection{IoT Sensor Node}\label{IoT}
To make the model more simple, it is assumed that there is one gateway  per farm as illustrated in Figure \ref{fig-01} (but there can be more than one in reality). The gateway processes data transmitted from  nano-sensors within its range which it sends to the mining nodes, while  also storing information from other mining nodes on the level of chemicals within the region. 

\subsubsection{Data Processing Operation}
The chemical-detecting indexes such as nitrogen (N), potassium (K), and phosphorus (P) have been defined as classes that consists of ranges. For example, the recommendation for  P and K indexes are between 3 and 4 in order to maintain optimum soil fertility level~\footnote{\url{https://www.teagasc.ie/crops/soil--soil-fertility/soil-analysis/}}. Our study categorizes $RF$ data collected by each gateway node from its nano-sensors into five classes termed Response Classes (RCs). RCs are defined by dividing the sensor's $RF$ range (i.e., 0-100\%) into five non-overlapping regions, as presented in Table \ref{tab1}. Therefore, they correspond to the fixed ranges of $[A]$ in the sensor active region. This is followed by computing the frequency of the five RCs in each gateway using the $RF$ values gathered over a period of time, resulting in Table \ref{tab2}. This table is referred to as a \textit{one-way frequency table} as it enables the derivation of conditional one-way relative frequency (i.e., marginal probability) by dividing each row by its sum of row frequency values. The relative frequency values are called conditional probabilities. That is, for instance, $P(1|A) = f_{A1}/f_A$ is the probability of the gateway node-1 being in the response class $A$. Algorithm \ref{Al2} summarizes the process of computing the relative frequencies for period $t$. We are, however, interested in the probability of a selected gateway (i.e., farm) being in a particular RC (i.e., $P(A|1)$- inverted value of $P(1|A)$). Therefore, these conditional probabilities are fed into the Bayesian updating method explained in the next section to compute those probabilities.

\begin{table}[!t]
  \begin{center}
    \caption{One-way response class (RC) frequency.}
    \vspace{-.25cm}
    \begin{tabular}{|c|cccc|c|} 
    \hline
      \backslashbox{RC}{Node}& 1 &  2 &  $\cdots$ & N  & Total\\
      \hline
       A & $f_{A,1}$ & $f_{A,2}$ &  $\cdots$ & $f_{A,N}$ & $f_A = \sum_{i =1}^N f_{A,i}$ \\
      $\vdots$     &  $\vdots$ &  $\vdots$ &   $\ddots$ & $\vdots$ & \vdots \\
      E & $f_{E,1}$ & $f_{E,2}$ &  $\cdots$ & $f_{E,N}$ & $f_E = \sum_{i =1}^N f_{E,i}$\\
      \hline 
    \end{tabular}
  \label{tab2}
  \end{center}
\end{table}

\begin{algorithm}[t]
\DontPrintSemicolon
\SetAlgoLined
\SetKwInOut{Input}{Input}\SetKwInOut{Output}{Output}
\Input{ $TN_{ID}$s  and \# of nano-sensors ($K$) }
\Output{$P( TN_{ID}| \Theta)$}
\BlankLine
\ForEach{$i \in \{N\}$}{    
    \BlankLine
    \ForEach{$j \leftarrow 1 \quad \text{to} \quad K$}{
        $\{RF_{j}\}_i =$ \textbf{Algorithm \ref{Al1}}\; 
    }
     $\left[f_i\right]_{5 \times 1} \leftarrow$ assign RC labels to $\{R_f\}_i$ and calculate their frequency\;
     collect $f_i$s into $ F_{5 \times N}$ \;
}
$P( TN_{ID}| \Theta) = F_{i \times N}/ \sum_{j = 1}^N F_{i, j} \quad \text{for} \quad  i = 1, \cdots, 5 $
\caption{Computing $P( TN_{ID}| \Theta)$ }
\label{Al2}
\end{algorithm}

\subsubsection{Sequential Bayesian Updating}\label{sbu}
In this section, Bayesian theory is briefly introduced and then the Sequential Bayesian Updating (SBU) method is discussed for computing the probability distribution of any selected gateway node (i.e., farm) being in the five RCs. The details given in section \ref{IoNT} and \ref{IoT} are used here to compute the probability distribution required for the SBU.   

Bayesian theory is commonly used in statistical inferences as it allows updating the inverted conditional probability based on the latest collected data/evidence. We assume there are $N$ gateway nodes and two random variables, which are the  selection of a gateway node ($\mathbb{G}$) and a RC ($\Theta$). For each of these random variables, their sample spaces are then $\mathbb{G} = \{ 1, 2, \cdots, N\}$ and $\Theta = \{RCs\}$. If a node $i \in \mathbb{G}$ is selected, the probability of the selected node being in the $j^{th} \in \Theta$ (RC) is computed using the Bayes theory as represented as follows:

\begin{equation}\label{Bay}
    P(j \in \Theta|i \in \mathbb{G}) = \displaystyle \frac{P(i \in \mathbb{G}|j \in \Theta)P(\Theta)}{P(\mathbb{G})},
\end{equation}
where $P(\Theta)$ is known as the prior probability distribution function (PDF) and represents the strength of the belief of a node being in the five RCs. The likelihood of the result given the prior distribution is represented by $P(i \in \mathbb{G}|j \in \Theta)$, where $P(\mathbb{G})$ is known as the evidence (or data) and computed as $\sum_{\forall j \in \Theta} P(i \in \mathbb{G}|j \in \Theta)$. Applying this formula for all gateway nodes, the probability of each node being in the five RCs  $[P(j|i)_{i = 1}^N]_{j = 1}^5$ can be computed to produce the matrix  $P(\Theta|\mathbb{G})_{5 \times N}$. This matrix is known as the posterior PDF computed based on the set of RC frequency data samples over a time period $T$. 

In the SBU, when a new RC frequency data sample is collected over the time period $(T+1)$, the new posterior PDF is computed by using $P(\Theta|\mathbb{G})_{5 \times N}$ as the prior probability matrix, as is represented as 
\begin{equation}
    P(\Theta|\mathbb{G})_{(T+1)} = \displaystyle \frac{ P(\mathbb{G}|\Theta)_{(T+1)} P(\Theta)_{(T)} }{P(\mathbb{G})_{(T+1)}}, 
\end{equation}
where  $P(\Theta)_{(T)} = \displaystyle \left[\Pi_{ k = 1}^{(T)}P(\mathbb{G}|\Theta)_k\right] P(\Theta)_0$. Based on this, the probability of the $i^{\text{th}}$ node being in the five RCs over the time period $(T+1)$ can be computed as
\begin{equation*}\label{psp}
    P(\Theta  | i )_{(T+1)} = \displaystyle \frac{P(i|\Theta )_{(T+1)} P(\Theta )_{(T)}}{P(i)_{(T+1)}}, 
\end{equation*}
where $P(\Theta )_{(T)} = \left[\Pi_{k = 1}^{(T)} P(i|\Theta)_k\right] P(\Theta)_0$. The  $P(\Theta|\mathbb{G})$ updating process will continue until there are changes in the posterior PDF that reaches a certain threshold. Algorithm \ref{Al3} summarizes the SBU steps. The optimal number of updating steps required for detecting the level of chemicals in a farm  will be discussed in detail in the next section.

\begin{algorithm}[!t]
\DontPrintSemicolon
\SetAlgoLined
\SetKwInOut{Input}{Input}\SetKwInOut{Output}{Output}
\Input{ \textbf{Algorithm \ref{Al2}} \& $P(\Theta)_{T-1}$}
\Output{$P( \Theta |TN_{ID})_T$ and $P(\Theta)_{T}$}
\BlankLine
Initialization: $\epsilon  = 1 \times 10^{-5}$\;
\While{$d \leq \epsilon $}{
$P( TN_{ID}| \Theta)_T \leftarrow$ execute Algorithm \ref{Al2} \tcc{relative frequency}
\ForEach{$ i \in \{TN_{ID}\} $  }{    
    \tcc{Posterior Probability, $PoS_{5 \times N}$}
    \BlankLine
    $P_i = f_i * p_i$,
    where $f_i \in [P( TN_{ID}| \Theta)_T]_{(5 \times i)}$ and $p_i \in [P(\Theta)_T]_{5 \times i}$ \;
    \eIf{$\sum P_i == 0 $}{
        $PoS_i = 0$ 
        }
    {
        $PoS_i = P_i / \sum P_i$
    }
    collect $P_i$s into a matrix $PoS_{5 \times N} $\;
    \BlankLine
    \tcc{Update Prior Probability, $PrI_{5 \times N}$}
    Indexes $k$ of $f_i$ where $f_i == 0$ \; 
    \If {number of $ k > 0 $}{
            Replace entries of $P_i$ at $k$ indexes by the values of $p_i$ such that $P_i[k] = p_i[k]$
        }
    \BlankLine
    collect $P_i$s into a matrix $PrI_{5 \times N}$\;
}
$P( \Theta |TN_{ID})_T = Pos$ and $P(\Theta)_{T} = PrI $\;
$d = \sum |P( \Theta |TN_{ID})_T - P( \Theta |TN_{ID})_{T-1}| $\;
$T = T + 1$\;
}
\caption{Computing the posterior probabilities using SBU}
\label{Al3}
\end{algorithm}

\subsection{Mining/Functional Nodes}
It is assumed that mining nodes are trustable entities and have the authority to control the BC. The mining nodes also use the Joint Cloud service for performing block mining \cite{30}. In the Joint Cloud service, the Cloud service providers may consist of government bodies such as an Agricultural Department Agency or authorized pharmaceutical companies. 

\subsubsection{Joint Cloud and incentive for mining}
The use of IoT sensors in SF is a business model for sensor Cloud. In this model, the Cloud provides IoT as a service by collecting and aggregating IoT-sensor data. The Cloud provider can place the sensors in the farms and collect the sensor data for the regulators ( miners in the present study) for verification and traceability. In this business model, the farms will outsource the IoT data collection and report the job to the IoT Cloud service. This IoT sensor as a service model can be implemented using BCs following the two steps given below. 
\begin{itemize}
    \item[1.] BC mining can be used as an incentive for proving the IoT sensor service. For example, in a PoW-based BC system, the IoT sensor service provider (i.e., miner) can gain  benefits by collecting mint tokens as it produces new blocks. 
    \item[2.] The IoT service is a geolocated service, where the quality of service depends on the location of the sensors and the cloud to reduce latency. The geolocated property supports joint cloud system. In this system, there will be multiple IoT cloud providers and they collaborate to provide a chemical traceability service for the regulators. In this interaction model between the service providers, BCs are very useful as IoT cloud providers that do not have trust between each other.  
\end{itemize}
The Clouds process the data from each farm (i.e., gateway) and offers certification through a colored token, which represents the level of chemical used on the farm. They also perform credit and tokens transactions between the farms and the regulators to offer (or charge) a certain amount of credits based on the quality of products and tracking the traceability, respectively. We will first describe the process of credit exchange and this will be followed by a description on the full functionality of the mining process of each nodes. 

\subsubsection{Farm credits}\label{crdit}
Let us assume that $Cr_T$ amount of credits is assigned by a governmental agency or miners to each farm when it joins the BC network at $T = 0$. For each mining step, the credibility is computed following the three steps given below.
\begin{itemize}
    \item [1.] Farmers are rewarded or penalized a certain amount of credits for either complying or not complying with the chemical standards.
    \item [2.] A farm is compliant with the chemical standards if it has a higher probability (i.e., at least of not being in the class $E$ (i.e., $P(\sim E) \geq .8$),
    \item [3.] The rate of change in the amount of remaining credits after several mining steps represents the credibility.
\end{itemize}
For any farm, the amount of credits at the $T^{\text{th}}$ mining step is computed as:
\begin{equation}\label{cr}
      Cr_{T} = \begin{cases} 
      Cr_{T-1} - P_{(T,E)} L +  P_{(T,\sim E)} M & f_{(T,E)} \neq f_{(T-1, E)}, \\
      Cr_{T-1} & \text{otherwise}, 
   \end{cases}
\end{equation}
where, $L = e^{\alpha f_{(T,E)}}, M = e^{\alpha f_{(T,\sim E)}}$ and  $f_{(T, E)}$ is the cumulative frequency of a farm being in the class $E$ up to the $T^{\text{th}}$ time step, $P_{(T, E)}$ is the probability of that farm being in the class $E$ at the $T^{\text{th}}$ mining step, and $\alpha \in \mathbb{R}$ is a constant and termed as the credibility tuning parameter. The amount of credits reduced or reward will exponentially increase if there is a change in the frequency of the response class $E$. That is, if $f_{(T,E)} \neq f_{(T-1, E)}$), then $P_{(T,E)} e^{\alpha f_{(T,E)}}$ amount of credits is reduced from the available credits $Cr_{T-1}$ of farms which belong to the class $E$,  while the amount of credits is added to the farm for not being in the class $E$ is $P_{(T,\sim E)} e^{\alpha f_{(T,\sim E)}}$. The amount of credits remains the same (i.e., $Cr_T = Cr_{T-1}$) if there is no change in frequency of being in the class $E$.  

\begin{algorithm*}[t!]
\DontPrintSemicolon
\SetAlgoLined
\SetKwInOut{Input}{Input}\SetKwInOut{Output}{Output}
\Input{parameters required for \textbf{Algorithm \ref{Al1},\ref{Al2}}, and \textbf{\ref{Al3}}, and $Tw$, $\alpha$}
\Output{block}
\BlankLine
 
Initialization:
 Genesis block containing $P(\Theta)_0$ and  $Cr_0$ \;
\BlankLine
\ForEach{$T \in$ \# of block mining steps}{    
    \tcc{ select a FN and generate its SK and PK}\;
    $FN_{k} \leftarrow$ select any id from $\{ FN_{id} \}$\; 
    ${SK_{(FN,k)}, PK_{(FN,k)}} \leftarrow $ D-H key exchange service\;
    \BlankLine
    \tcc{Extract the prior probability from the latest block}
    \eIf{ T = 0}{
     $P(\Theta)_{T-1} \leftarrow $ from the genesis block}
    {
    $P(\Theta)_{T-1} \leftarrow $ from the last block of the BC of the $FN_{k}$'s ledger\;
    }
    \tcc{Collecting and sharing data at TNs}\;   
    $\forall i \in \{TN_{id}\}_{i=1}^N$\;
        $ \hspace{1cm} {SK_{(TN,i)}, PK_{(TN,i)}} \leftarrow $ D-H key exchange service\;
        $\hspace{1cm}  SSk_{(TN,i)} \leftarrow $ D-H key service$(SK_{(i)}, PK_{(FN,k)})$,\;
       $\hspace{1cm} $ collect RC frequency $\{F_{i}\}_{t = 1}^Tw \leftarrow$ execute \textbf{Algorithm \ref{Al2}}, \;
       $ \hspace{1cm}  a_{i, T}\leftarrow $ AES-GCM encryption($\{F_{i}\}_{t = 1}^Tw, SSk_{i}$), \;
        $ \hspace{1cm}  FN_k$ collects $Data_{(i,T)} = [ a_{(i,T)}, PK_{(TN,i)} ]$.\;
    
    \BlankLine
    \tcc{Processing data and block mining at FN} 
    $\forall i \in \{TN_{id}\}_{i=1}^N$, \;
        $\hspace{1cm}  PK_{(TN,i)} \leftarrow $ get from $data{(i,T)}$,\;
        $\hspace{1cm}  SSk_{(FN,k_i)}  \leftarrow $ D-H key service$(SK_{(FN,k)}, PK_{(TN,i)})$,\;
        $\hspace{1cm}  b_{(i,T)} \leftarrow$ AES-GCM decryption$(a_{(i,T)}, SSk_{(FN,k_i)})$, \;
        $\hspace{1cm}  \text{collect} b_{(i,T)} \text{to} F_{t}$.\;
      
    $P(\Theta|TN_{ID})_T \leftarrow$  execute \textbf{Algorithm \ref{Al3}} \;
    \BlankLine
    
    \ForEach{ $i \in \{TN_{id}\}_{i = 1}^N$}{
    Generate color token ($CT_{(i,T)}$) $ \in \leftarrow  [P(\Theta|TN_{ID})_T]_{5 \times i}$\;
    Update $[f_T]_{(T,E)}$\;
    Credits ($Cr_{(i,T)}$) $\leftarrow Cr_{(i,T-1)} + \left[P(\Theta|i)_{(T,E)} e^{\alpha [f_i]_{(T,E)}}\right] + \left[ (1 -P(\Theta|i)_{(i,\sim E)}) e^{\alpha [f_i]_{(T,\sim E)}}\right]  $\;
    collect $ \left[CT_{(i,T)}, Cr_{(i,T)}\right]$ into $Data_{(T)} $\;
    \BlankLine
    \tcc{mine TN block}
    $Data_{(i,T)} \leftarrow$ AES-GCM encryption( $\left[CT_{(i,T)}, Cr_{(i,T)}\right]  , SSk_{(FN,k_i)}$)\;
    mine block for TN network adding $[D_{(i,T)}, PK_{(FN,k)}]$\;
    }
    \tcc{mine FN block}
    mine block for FN network adding $[D_{(T)}]$
    \BlankLine
    \tcc{Updating ledgers}
    $\forall i \in \{TN_{ID}\}$ and $\forall k \in \{FN_{ID}\}$, update ledgers
    
    $tn \in TN_{ID}, SSk_{tn} \leftarrow key_{tn}(PK_{FN})$\;
}
\caption{Block mining}
\label{Al4}
\end{algorithm*} 

\subsubsection{ Block Chain Network}\label{bc_farm}
This section presents how the BC  is incorporated with the SBU, joint Cloud system, and the credit computing approach for detecting the level of chemicals used on the farms. The BC network employed here is a private BC. 

The gateway nodes and the mining nodes in the BC network are termed as the transaction nodes (TN) and functional nodes (FN), respectively. The TNs collect data and send them to a selected FN which aggregates the TNs' data and mines two blocks for the TN and FN networks (more details are given below). Furthermore,  data communicated between the TN and FN networks are encrypted and then sent following a key exchanging mechanism similar to peer-to-peer key agreement method proposed in order to protect data privacy, security, and integrity. The functionality of the BC network uses three steps; data sharing, data processing and block mining, and BC updating, and each of these steps are described as follows  (Algorithm \ref{Al4} summarizes these steps).

\begin{itemize}
    \item [1.] \textbf{Data sharing:} Algorithm \ref{Al2} is executed at each TN for collecting RC frequency data samples for a period of time $T$ (this is termed as data stream). Each data stream is then encrypted and access permissions are granted by using the authenticated symmetric encryption under the \textit{Advanced Encryption Standard Galois Mode} (AES-GCM)\footnote{\url{https://cryptography.io/en/latest/hazmat/primitives/aead/}} and using the  Shared Secret Key (SSk). The encrypted data streams are then submitted to a selected FN, which is the miner.
    
    The AES-GCM enables decryption of data by using the same SSk key that is used for encrypting the data. This encryption generates compressed data in plain-text which contains a key value used for integrity protection and authentication at the decryption. Hence, any FN that contains the encrypted key can verify the integrity of the encrypted data and perform an authenticated decryption. Each TN will share the encrypted data and the its Public Key (PK) with a selected FN.   

    The PK and secret key (SK) for each TN is generated using the \textit{Diffie-Hellman} (D-H) key exchange service, which facilitates the sharing of a common secret key between two or more parties. The SSk at a TN is generated by using its SK and PK of the FN through the D-H service. This is one of the reliable techniques for sharing data between unknown parties as the D-H service avoids sending the SK away from its owner while also allowing the owner to revoke sharing data at any time \footnote{\url{https://cryptography.io/en/latest/hazmat/primitives/asymmetric/dh/}}.

   \item[2.] \textbf{Data processing and block mining:} Prior to creating a block at the FN, the data are processed as follows:
   \begin{itemize}
       \item [a.] Data coming from a TN is decrypted through the SSk generated using the FN's SK and the PK of the TN included in the data.
       \item [b.] A ML task is executed as a smart contract  to derive insights about the level of a chemical (this ML task is further described below).
       \item [c.] By using the insights derived in the previous step, a color token is generated to represent the level of a chemical used. The color token represents the variability in chemical levels over a period of time $T$.  
       \item [d.] Based on the color token generated for each farm, the amount of credits held by each farm is updated.
   \end{itemize}

    {\bf Smart contract/ML task:} The SBU updating process presented in Algorithm \ref{Al3} is executed to derive the updated posterior PDF. Then for each TN, a color token is derived from its posterior PDF. A color token consists of five unique colors that correspond to the five RCs. The region occupied by each color in the color token is proportional to the probability of the level of a chemical being in the five RCs (explained in detail in the next section).  

    Since the TNs and FNs play two different roles in this system, the information required to be stored in the TN and FN networks is different. Therefore, two blocks are created for the FN and TN BC networks.  The block created for the FN network includes color tokens and credit transactions of all TNs because they are required to perform future block mining. When creating a block for the TN network, the color token and credit value of each TN are  encrypted together as they are sent away from the FN network. The AES-GCM and H-D methods explained above are used for the encryption. The new block is then created, including the encrypted data and FN's PK.

    Finally, the FN block is added to the FN ledger while the TN block is sent to at least one TN to add it to the TN network. The PoW technique \cite{18} is used for validation of the new blocks and then adding blocks into the BC ledgers. 
    
    \item[3.] \textbf{Updating BC:} After the TN's and FN's blocks are added to their BCs, all FNs and TNs update their ledgers accordingly. Any TN can derive the SSk by using its SK and FN's PK that is contained in the BC, which is also used for decrypting the data to have up-to-date information about the current level of chemicals used on the farm. Similarly, any FN can be a future miner as all necessary information required to perform a new mining task is contained in their BC. At the moment, the miner is selected randomly.
\end{itemize}
 
\subsection{Performance metrics}
Two performance metrics, \textit{Mean Squared Error} (MSE) and \textit{Accuracy} (AC), are used for assessing the performance of the BC-IoNT system. While MSE is used as the performance measure to decide how accurately the proposed BC system can detect the level of chemicals, the AC is the percentage of the number of TNs which the RC has identified correctly. These two metrics are computed as follows:
\begin{eqnarray*}
    MSE &=& \frac{\sum_{i = 1}^N(y_i - \hat{y}_i)^2}{N}   \\
    AC &=& \frac{\sum_{i = 1}^N f_i}{N} \times 100, \quad  f_i = 1 \quad  \text{ iff} \quad  |y_i - \hat{y}_i| = 0, 
\end{eqnarray*}
where $y$ and $\hat{y}$ stand for the actual and the predicted RC of which TN belongs to, respectively, and $N$ is the number of TNs (i.e., farms).


\section{Results} \label{sec4}
This section will evaluate the use of the proposed BC system for detecting the levels of chemicals in farmlands. First, experimental setups used for simulations are  briefly explained.  Next, the process of computing the $RF$ values from the nano-sensor signals is discussed. The variability in the probability of each farm categorized in the five RCs is explained as the third step, and this is followed by discussion on selecting the optimal parameters for generating color tokens effectively.  Finally, the color token is used to represent the levels of chemicals.
 
\subsection{Experimental setups}
It is assumed that the area covered by each gateway (TN) is the same and corresponds to a farm. However, in reality, a farm could contain a collection of such devices. The experimental procedure is as follows:

    \begin{table}[!t]
      \begin{center}
        \caption{Simulation Parameters}
        \begin{tabular}{ll} 
        \hline
           Parameter & Value \\
          \hline
          Association rate ($k_a$)              & $ 10^{-2} (MS^{-1})$  \\ 
          Disassociation rate ($k_d$)           & $ 10^{-3} (S^{-1})$     \\  
          Receptor concentration ($R_{max}$)    & 100 ($Ru$) \\
          Number of Farms ($N$)                 & 40   \\  
          gateways (node) per field           & 1  \\   
          Nano-sensors per node ($K$)           & 100  \\ 
          Analyte concentration  ($[A]$)        & $[\mathcal{U}(0,50)]_{1 \times N}$ \\ 
          Prior probability ($[P(\Theta)_0]_{(5 \times 1)}$)  & $ [0.2]_{(1 \times 5)}$\\
          IoNT signal threshold ($\varepsilon$) & $10^{-5}$\\
          Initial credits ($Cr_0$)              & 500 \\
          \hline
        \end{tabular}
         \label{tab3}
      \end{center}
    \end{table}

\begin{figure*}[!t]
  \begin{subfigure}[b]{.24\linewidth}
    \includegraphics[width=1\columnwidth]{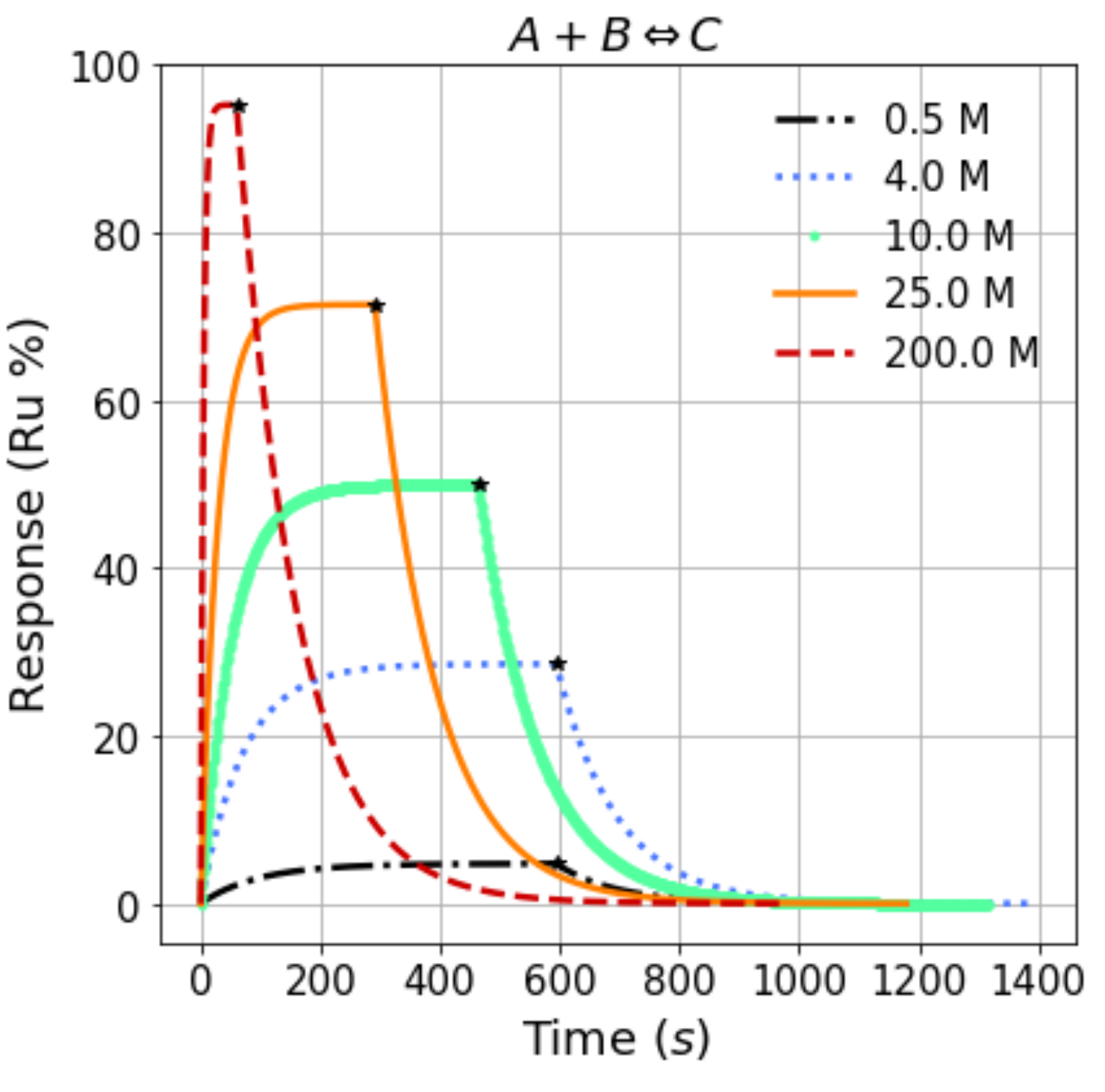}
    \caption{}
    \label{fig-02-a}
  \end{subfigure}
  \hfill 
  \begin{subfigure}[b]{.24\linewidth}
    \includegraphics[width=1\columnwidth]{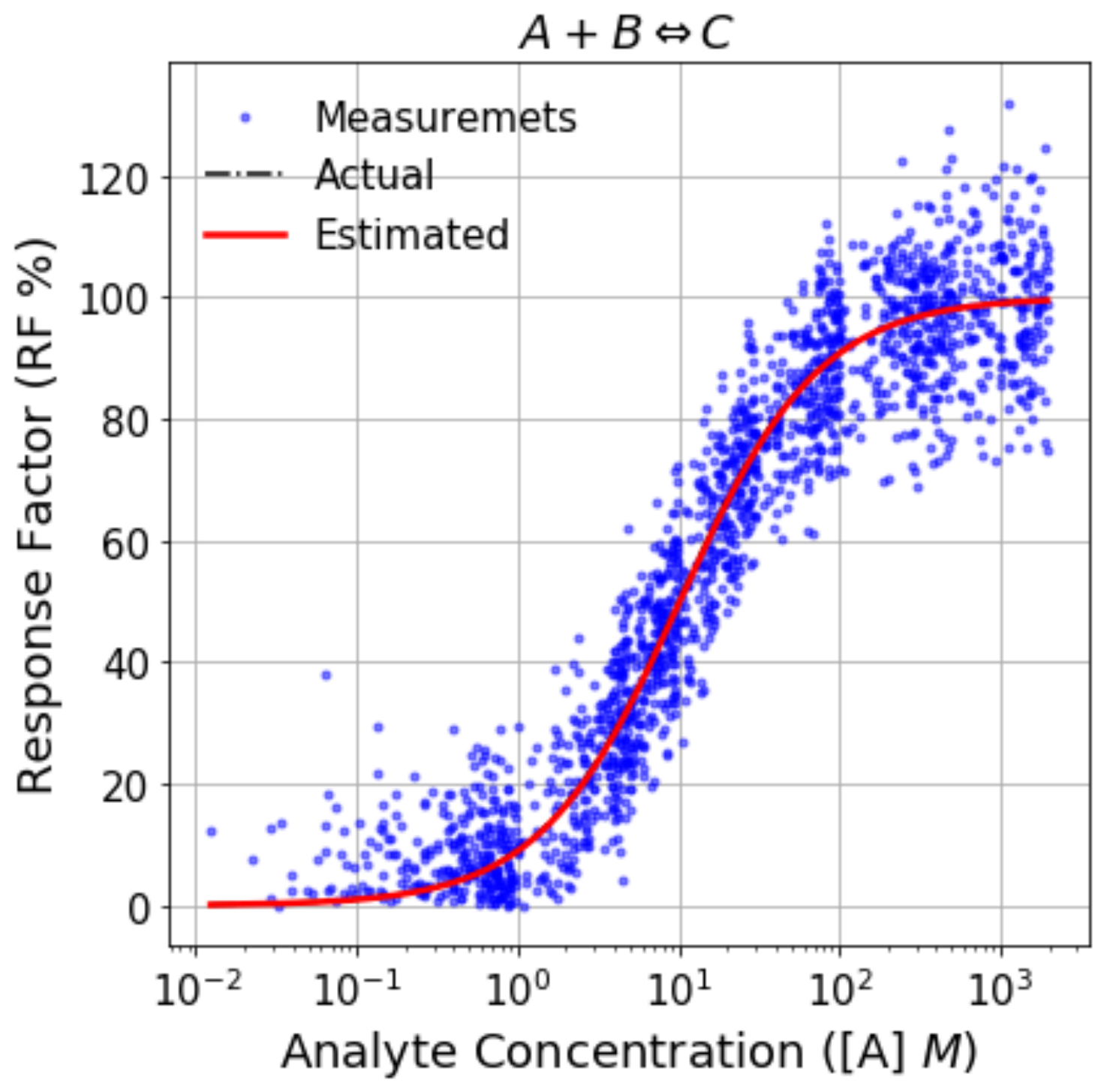}
    \vspace{-1.5em}
    \caption{}
    \label{fig-02-a1}
  \end{subfigure}
  \hfill 
  \begin{subfigure}[b]{.5\linewidth}
    \includegraphics[width=\columnwidth, height = 3.5cm]{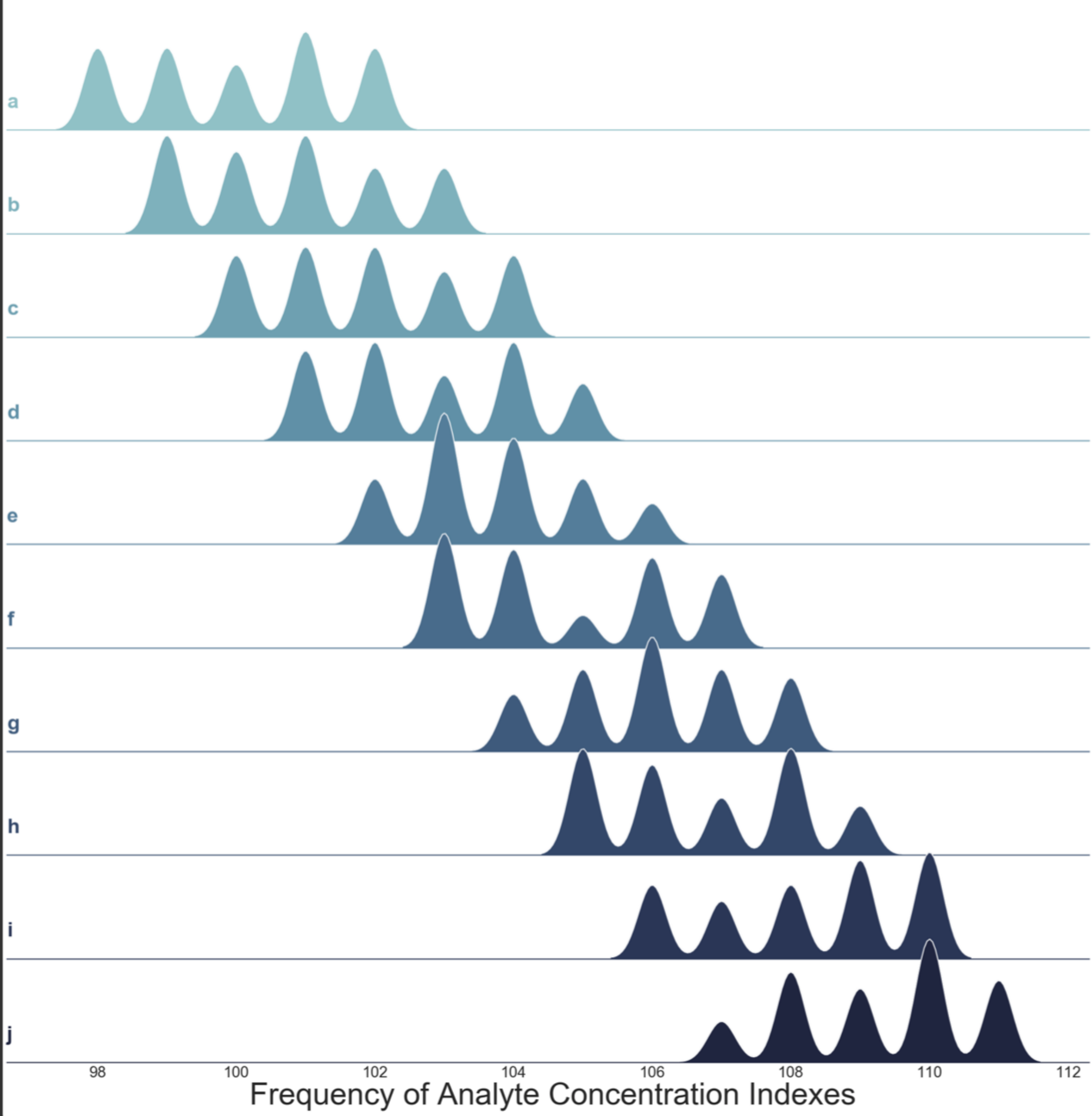}
    \vspace{-1.5em}
    \caption{}
    \label{fig-02-b}
  \end{subfigure}
  \vspace{-1.0em}
  \caption{Nano-sensor response, (a) for different $[A]$ values, (b) for a range of $[A]$ values (i.e., response Factor model), and (c) frequency of RCs over a period of time. }
  \label{fig-02}
\end{figure*}

\begin{itemize}
    \item [1.] We assume that a chemical A is applied over the set of farms at different concentrations and this range is presented as $\mathcal{U}(0, 50)$. The reason for selecting this range is based on the $k_a$ and $k_d$ values and explained in detail in section \ref{rs1}. The distribution of $[A]$ on each farm was also varied by using the Gaussian distribution in response to different field conditions.
   
    \item [2.] For a given period ($T$), the TNs collect a set of RC frequency data samples and then pass this to the FN, where a smart contract is executed to detect the levels of chemicals on the farms (i.e., $[A]$). The optimal number of RC frequency data samples required for deriving the level of chemicals in each farm with $p\leq 0.001$ accuracy is discussed later.
    
    \item [3.] The last step is to perform credit transactions for each farm where two blocks are created.
\end{itemize}

These three steps are repeated for each application of chemicals. All studies were based on simulated data generated by using the parameters presented in Table \ref{tab3}. In the simulation, these parameters are used, unless mentioned otherwise based on the evaluation. 
  
\subsection{IoNT sensor response and response factor}\label{rs1}
Algorithm \ref{Al1} was executed for different $[A]$ values to study the variability in the sensor response ($R$) over time. Figure \ref{fig-02-a} presents the variability in the $R$ for different $[A]$ values. With increasing $[A]$, the maximum $R$ increases, while the time taken to achieve the maximum $R$ reduces. In particular, when $[A] = 10 M$ (i.e., $k_D = [A]$), the maximum sensor response is $50 \%$ and thus, it proves the theoretical fact mentioned in \ref{rf}, which states that $k_D = [A]$, the $R_{max} = 50 \%$ (i.e., half of the receptors are occupied by the molecules of A). 

Obtaining information on the sensor active region is important as it provides prior insights about the sensor capabilities. Hence, the sensor active region was derived using two steps; (1) $RF$ values were derived for a set of $[A]$ values in the range $[10^{-2}, 10^3]$ and then randomized by adding $\mathcal{N}(0,1)$ error terms, and (2) a non-linear regression model was fitted to the RF values as the theoretical $RF$ model in (\ref{rf}) is a non-linear function. The fitted $RF$ model in Figure \ref{fig-02-a1} indicates that the $RF$ indicates a significant change in response to the change in $[A]$ within the range $10^{-1} \leq [A] \leq 10^{2}$. Therefore, this range was selected as the sensor active region and used to set up experiments for executing the proposed BC system. As shown in Figure \ref{fig-02-a1}, the concentration ranges covered by the RCs from $A$ to $E$ are increasing. Hence, the average maximum level of $[A]$ was set as $50M$, aiming to obtain a fair distribution of RCs over the farms. Otherwise, the distribution of the RC could be concentrated towards the RCs corresponding to higher RCs (e. g., $D$ and $E$) for the wider range of $[A]$.  

 \begin{figure}[t]
    \centering
    \begin{subfigure}[b]{0.5\linewidth}
        \centering
        \includegraphics[width=\columnwidth]{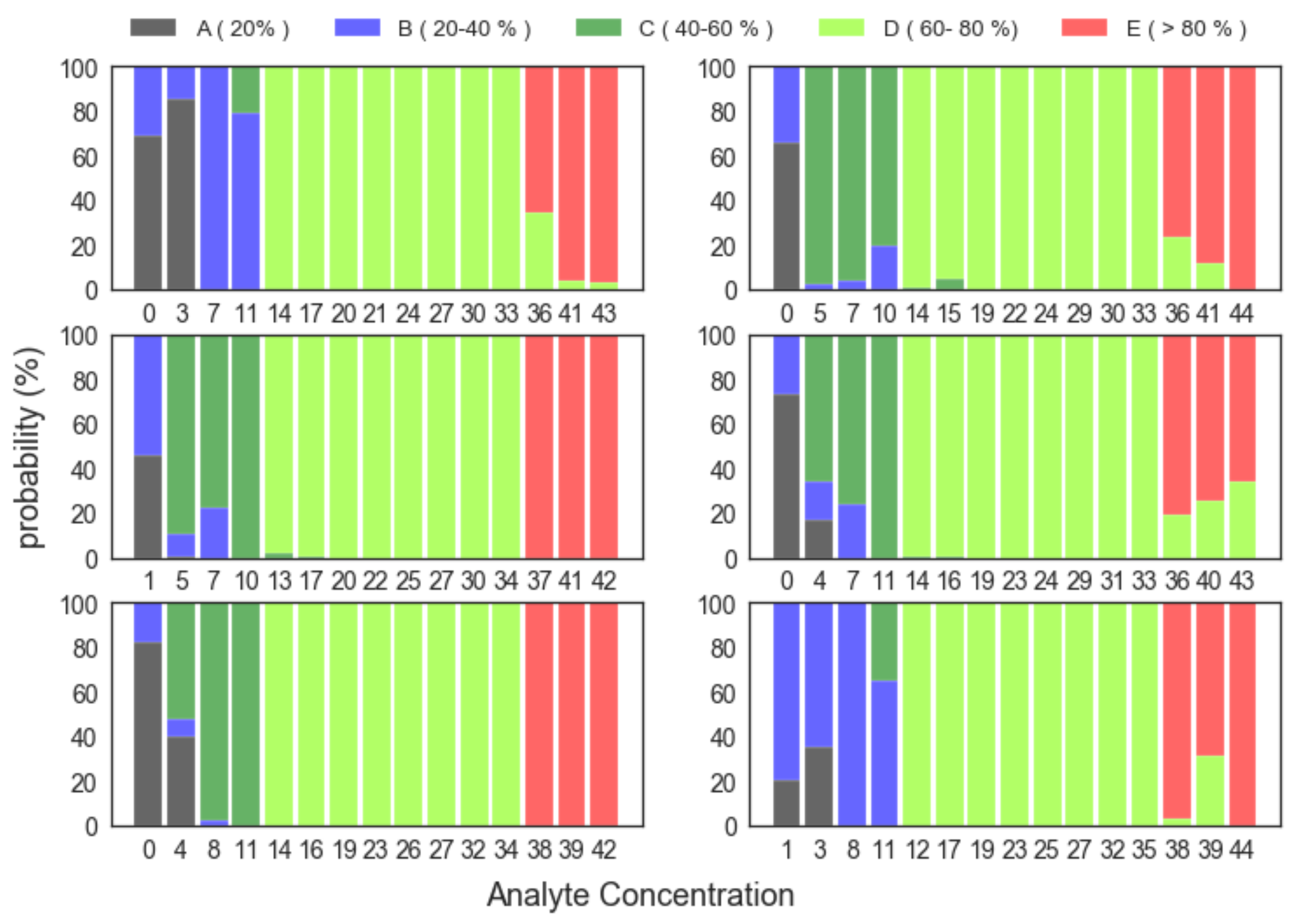}
        \vspace{-2.0em}
        \caption{}
        \label{fig-03-a}
    \end{subfigure}%
    ~ 
    \begin{subfigure}[b]{0.5\linewidth}
        \centering  
        \includegraphics[ width=\columnwidth]{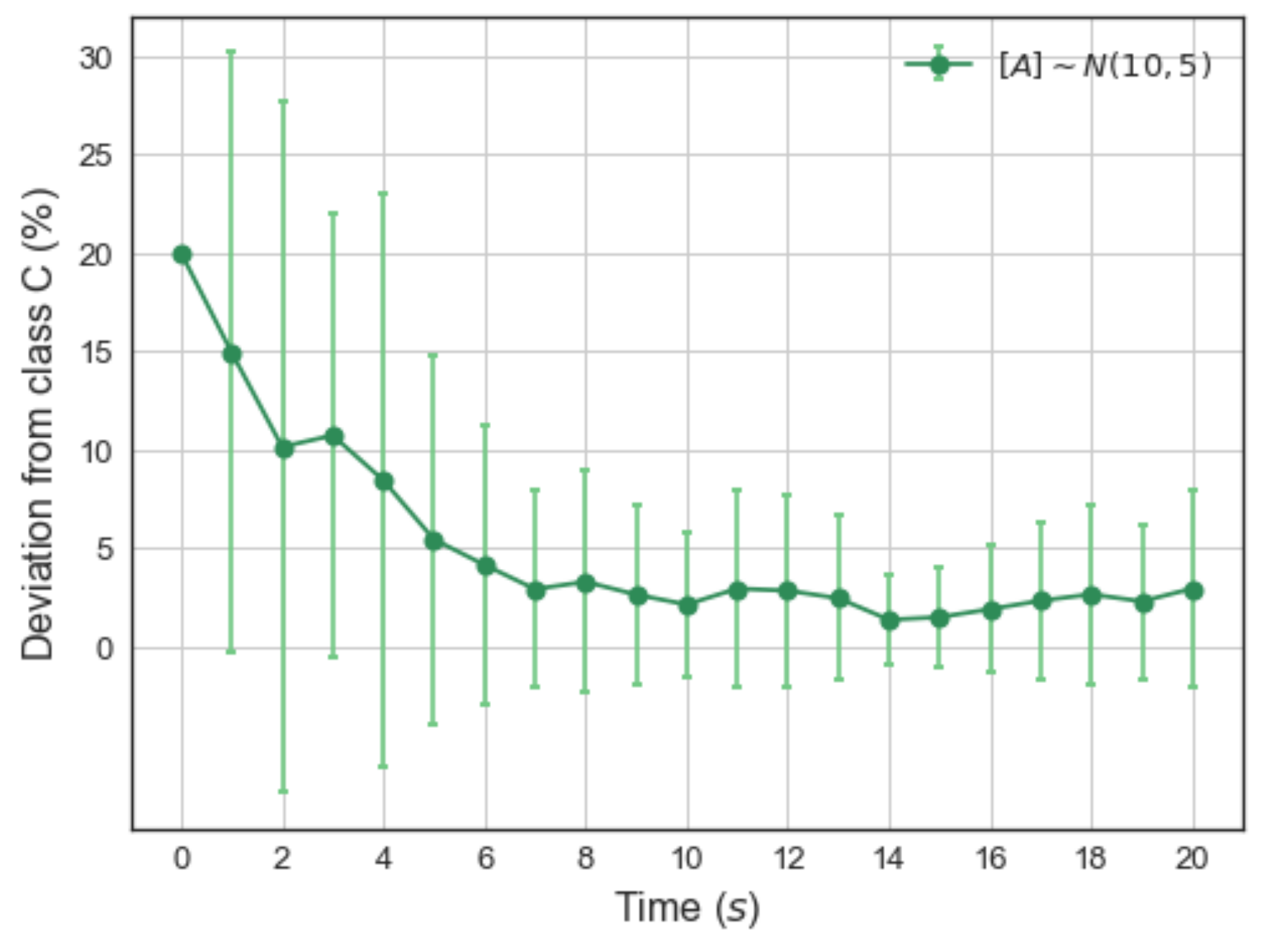}
        \vspace{-2.0em}
        \caption{}
        \label{fig-03-b}
    \end{subfigure}
    \vspace{-1.5em}
    \caption{Sequential Bayesian updating outcomes (i.e., probability) for detecting level of $A$, (a) when $[A]$ varied over the range $[0, 50]$ , where $[A] =0$ means $0 < [A] < 1 $ and (b) deviation from the class $C$.}
    \label{fig-03}
\end{figure}

\subsubsection{ Response Class Frequency }\label{frq}
To illustrate the variability in the RC frequency data collected in Table \ref{tab2}, Algorithm \ref{Al2} was executed at TNs for a period of time. Figure \ref{fig-02-b} shows the RC frequency distributions derived by using the kernel density estimation technique for ten TNs. The five peaks from left to right in each graph corresponds to the RCs from $A$ to $E$. Based on the height of the peaks, the RC of each farm can be determined.  However, the RC frequency distribution is a static measure and the level of chemicals could vary dynamically due to the influence of several time-variant factors such as land usage, weather, and prior chemical usage. Therefore, using only the RC frequency distribution is not sufficient enough to describe the availability of a certain chemical that has been used.  Here, the next section discusses how the current and prior level of a chemical can be incorporated for dynamically updating the current levels of a chemical by using the SBU approach integrated with BC technology.

\subsection{ Probability of deviation from the optimal response class} \label{dv}
This section first discusses the variability in the probability of the chemical levels on the farms and its position in the five RCs. This is achieved by randomly varying $[A]$ over the range $[0,50]$ and then the probability of deviation from the optimal RC. The optimal RC was selected as the class $C$. In reality, this could be any of the RCs based on the application requirements. 

To compute the posterior PDF, the SBU method in Algorithm \ref{Al3} was executed for several iterations, assuming that initially the level of a chemical in every farm has an equal chance of being in any of the five RCs (i.e., $P(\Theta) = [0.2]_{1 \times  5}$). Figure \ref{fig-03-a}, for instance, illustrates the variability in posterior PDF of $[A]$ in six farms that are within the five RCs with increasing $[A]$ for 15 probability updating steps. The probability of $[A]$ in farms being in the $D$ and $E$ RCs is greater for larger values of $[A]$ (at least $ \geq 15$), but when $[A]$ is below or around the equilibrium concentration (i.e., $[A] =  10 M$), the farms belong to the $A, B,$ and $C$ RCs. Similarly, the same procedure was repeated for a number of iterations by taking into account the variability in $[A]$ of forty farms and computing the probability of deviation from the class $C$ as the probability of not being in $C$ in each updating step. Figure \ref{fig-03-b} shows that  the deviation in probability reduces with increasing updating steps. Thus, by performing several probability updating steps, it can easily be recognized to which RC a selected farm’s chemical level is converging, thereby identifying the farms which are not compliant with the optimal chemical standards.

\begin{figure}[!t]
  \begin{subfigure}[b]{.48\linewidth}
    \includegraphics[width=\columnwidth]{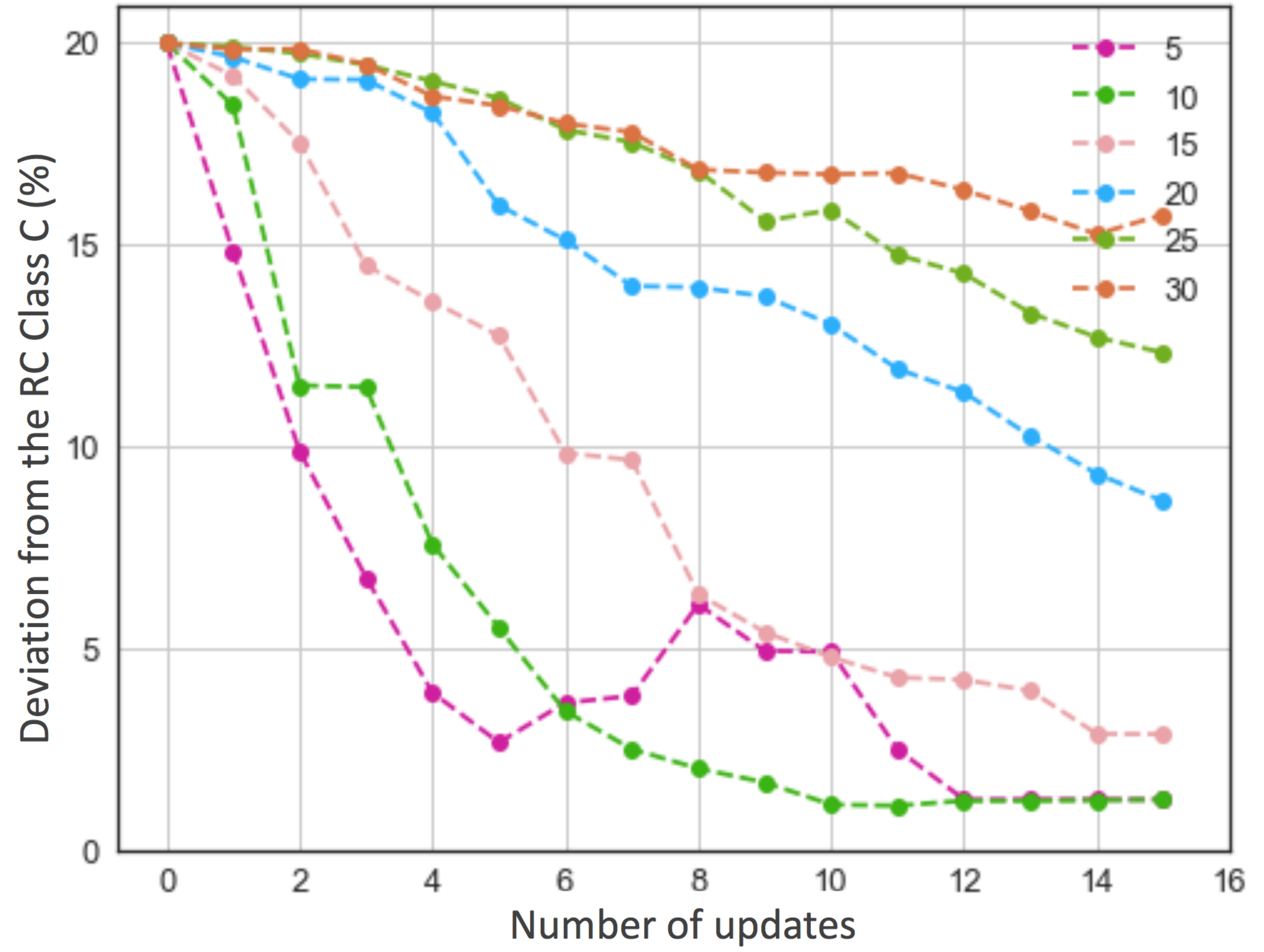}
    \vspace{-1.5em}
    \caption{}
    \label{fig-04-a}
  \end{subfigure}
  \hfill 
  \begin{subfigure}[b]{.5\linewidth}
    \includegraphics[width=\columnwidth]{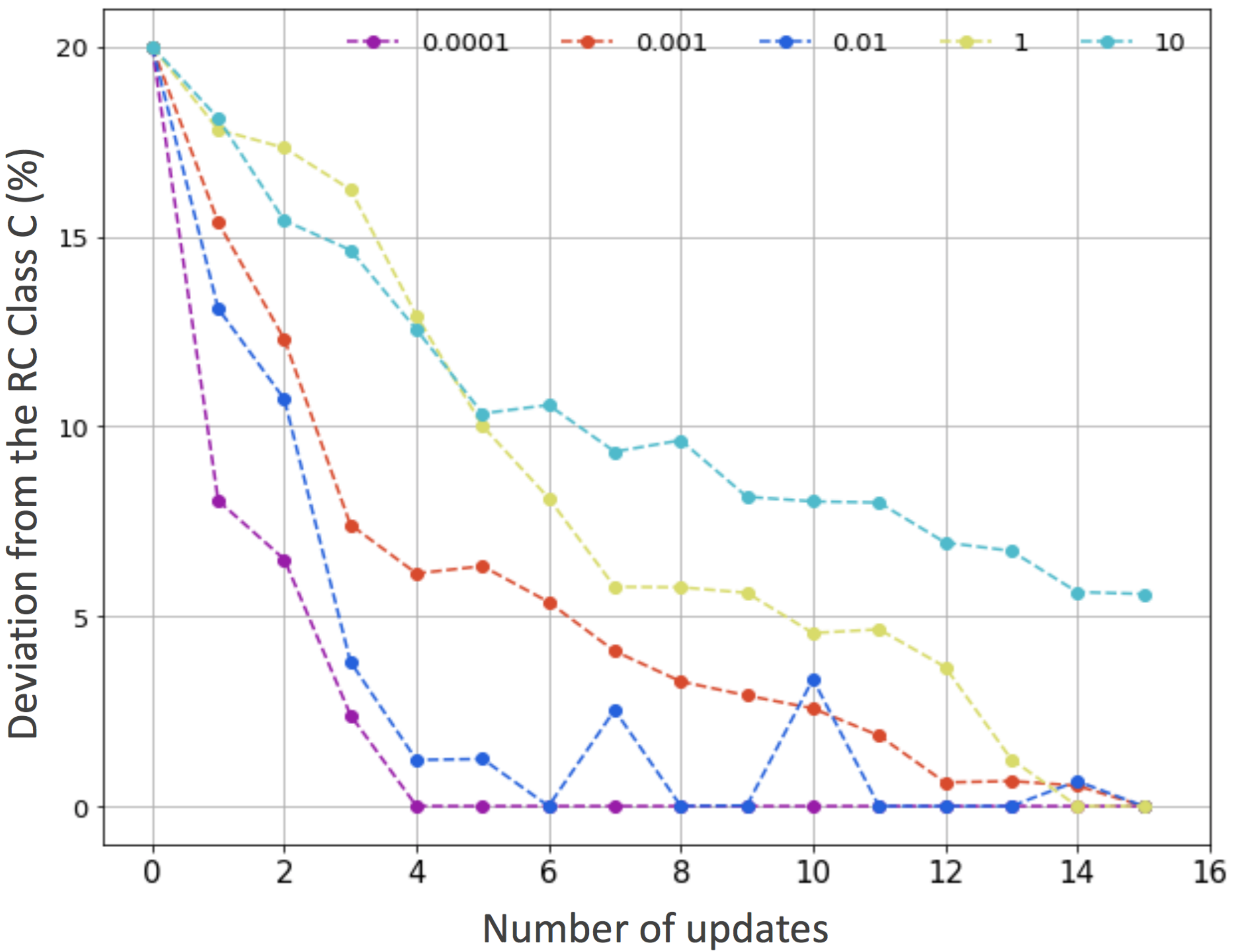}
    \vspace{-1.5em}
    \caption{}
    \label{fig-04-b}
  \end{subfigure}
  \vspace{-1.5em}
  \caption{The convergence of probability of deviation from the class $C$ for different, (a)  variability in levels of $[A]$ and (b) $k_D$ values. }
  \label{fig-04}
\end{figure}

However, this convergence rate varies depending on various factors. Two of the most important  factors are variability in $[A]$ within as well as among other farms and the affinity constant $k_D$. To illustrate their impact, the variability in the convergence rate of posterior PDF was explored for different variability levels in $[A]$ as well as a set of $k_D$ values. The outcomes for this analysis are presented in Figure \ref{fig-04}. The convergence rate was faster for smaller variability levels in $[A]$ within farms as well as smaller $k_D$ values (i.e., larger association rates). Therefore, these outcomes unveil  the criticality of deciding the optimal number of probability updating steps required for precisely deriving the RC of each farm.

\subsection{Optimal probability updating steps (sample frequency)}
This section discusses selection of the optimal number of probability updating steps. That is, the number of samples (say $S_{opt}$) that each TN is required to send to FN based on the variability in $[A]$ within, as well as among other farms, and this includes the optimal sample frequency (i. e., optimal value $T$ value termed as the time-window size ($Tw$)). 

According to Figure \ref{fig-02-a}, the time taken to reach the maximum response varies with the variability in $[A]$. Waiting until all nano-sensor nodes reach their maximum response to collect sensor responses could create extended latency in detecting the level of $[A]$. This issue can be mitigated by collecting sensor responses within a fixed period of time (i.e., $T_w$). However, too small a value of $T_w$ increases the $S_{opt}$, while larger $T_w$ values extends the time for collecting RC frequency samples. Therefore, the derivation of optimal values for $Tw$ and $S_{opt}$ was conducted in two steps as follows:

    \begin{figure}[!t]
        \centering
        \begin{subfigure}[b]{.48\linewidth}
            \includegraphics[width=.9\columnwidth]{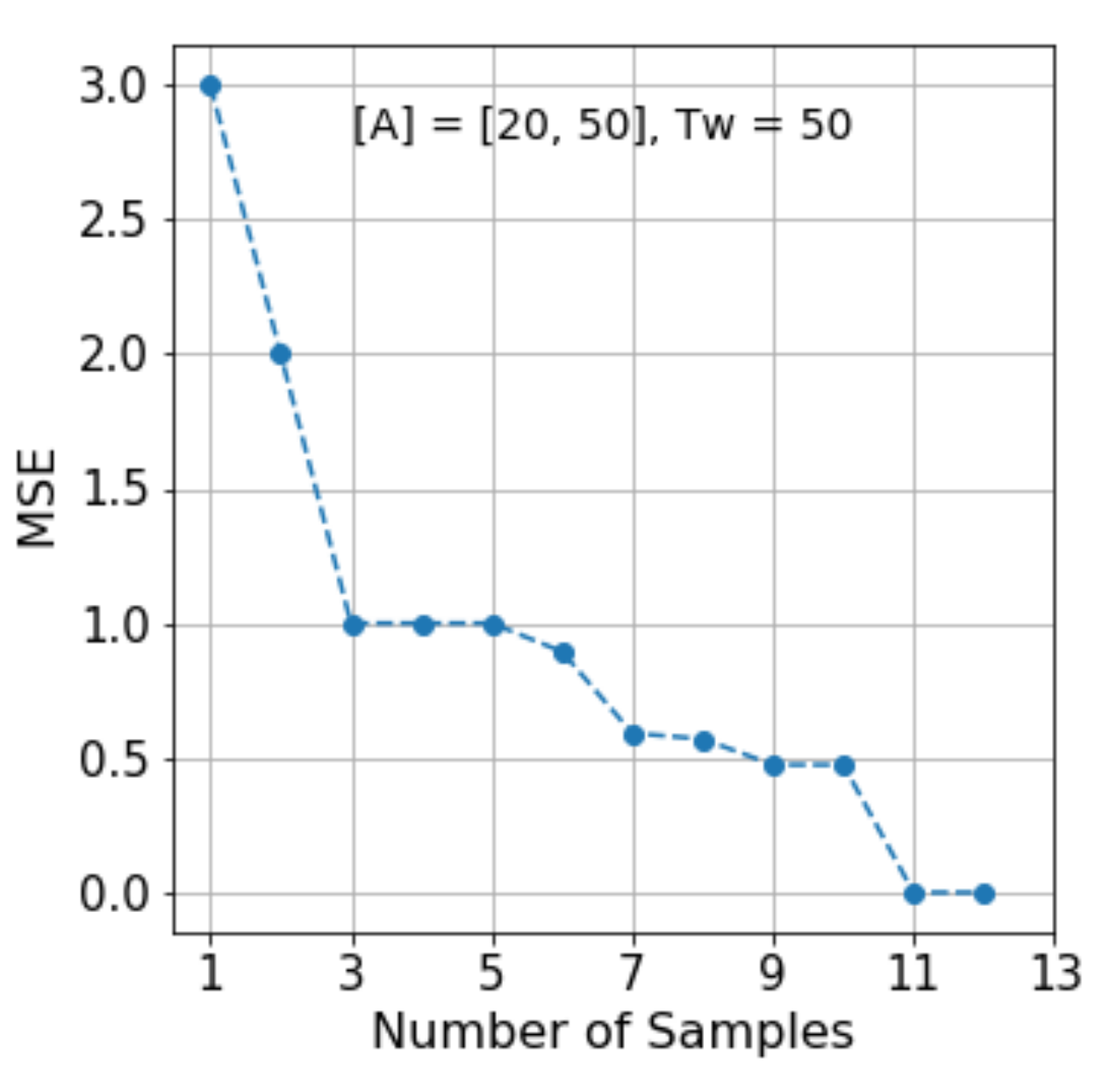}
            \vspace{-1.0em}
            \caption{}
            \label{fig-05-a}
        \end{subfigure}
        \hfill 
        \begin{subfigure}[b]{.48\linewidth}
            \includegraphics[width=.9\columnwidth]{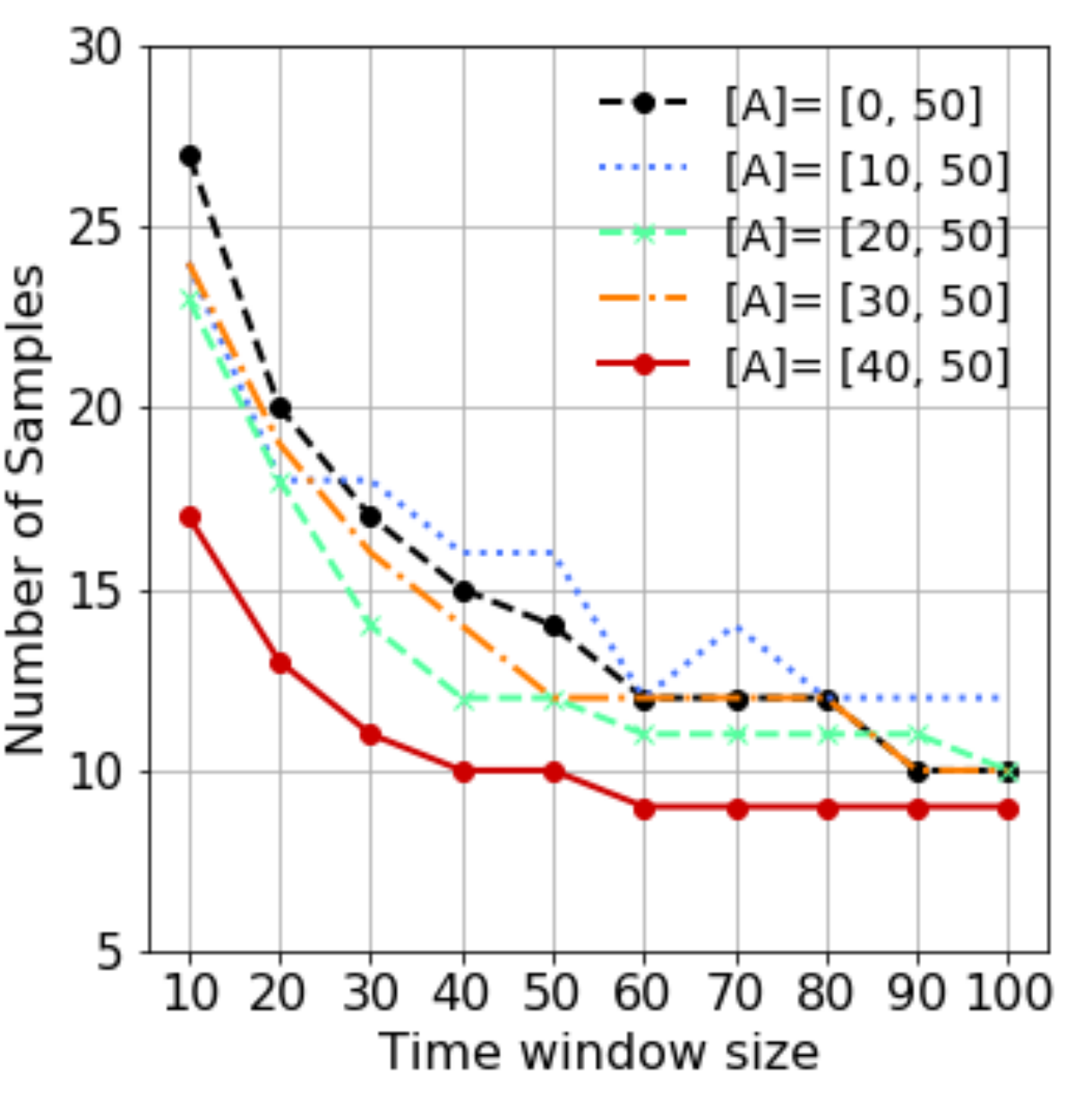}
            \vspace{-1.0em}
            \caption{}
            \label{fig-05-b}
        \end{subfigure}
        \vspace{-1.5em}
        \caption{ Selecting optimal number of samples required for detecting the level of $[A]$, (a) based on the variability in MSE and (b) with the increasing $Tw$ and range of inter-farm $[A]$.}
        \label{fig-05}
    \end{figure}
    
        \begin{figure*}[!t]
          \begin{subfigure}[b]{.32\linewidth}
            \includegraphics[width=.9\columnwidth]{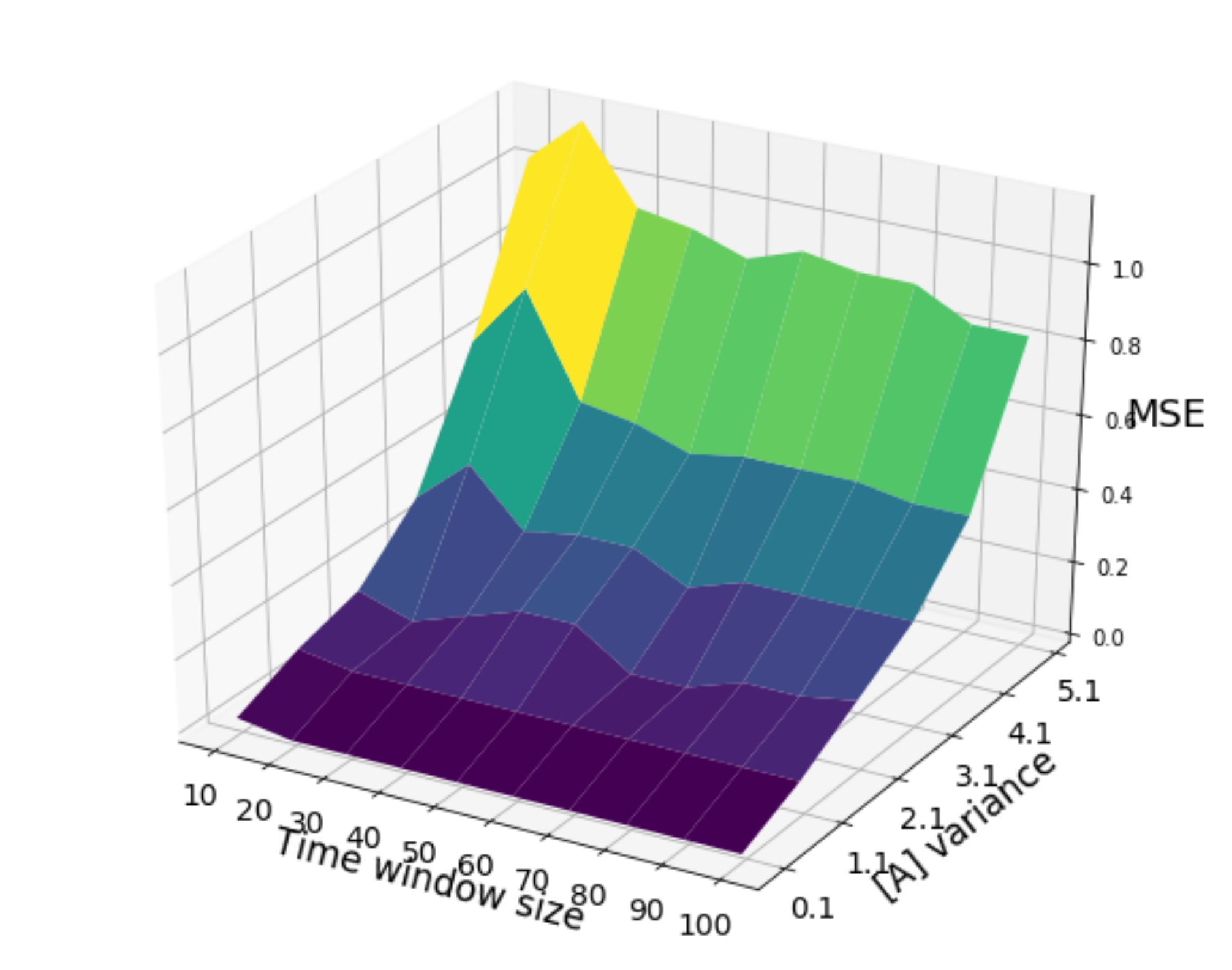}
            \vspace{-1.0em}
            \caption{}
            \label{fig-06-a}
          \end{subfigure}
          \hfill 
          \begin{subfigure}[b]{.32\linewidth}
            \includegraphics[width=.9\columnwidth]{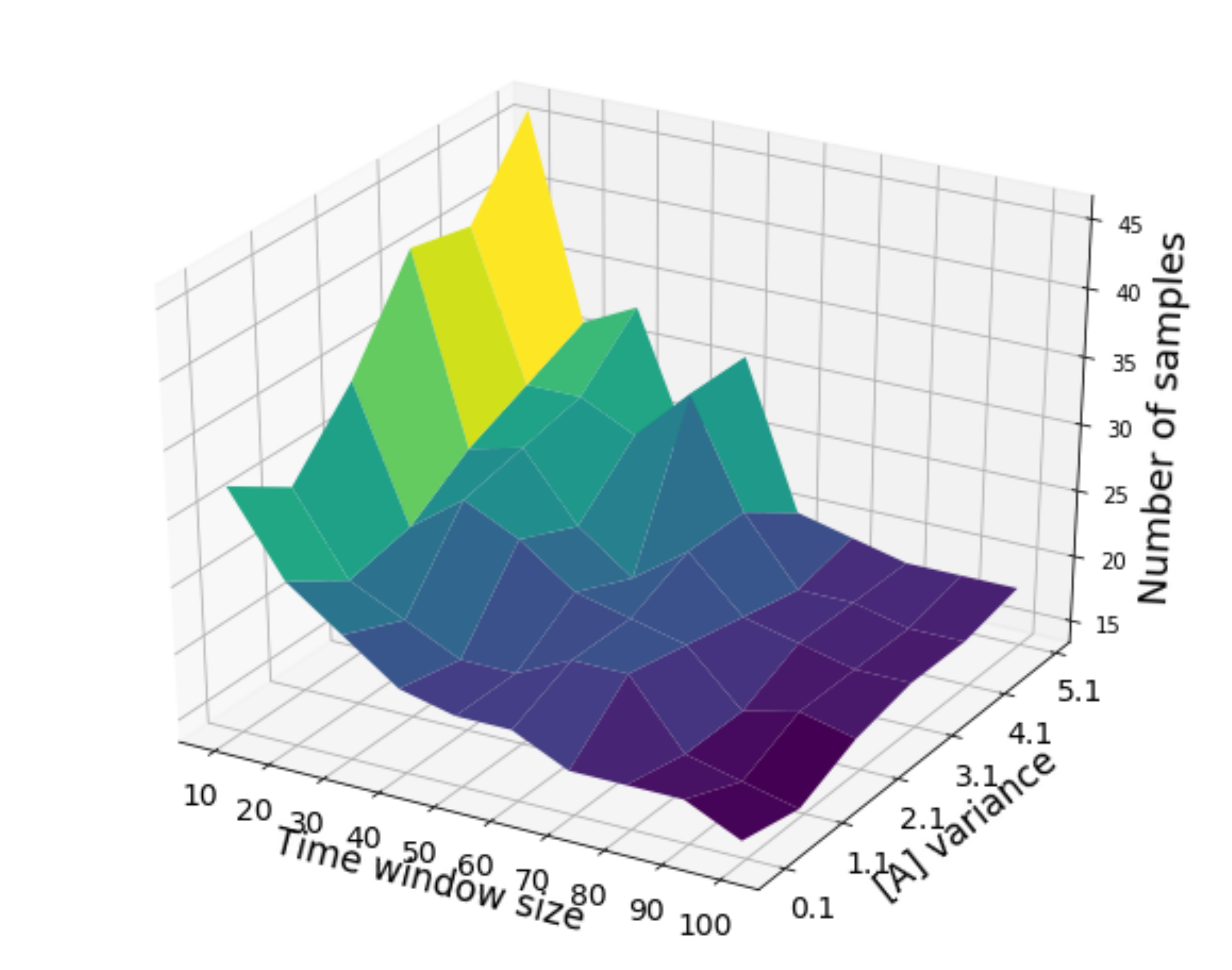}
            \vspace{-1.0em}
            \caption{}
            \label{fig-06-b}
          \end{subfigure}
          \hfill 
          \begin{subfigure}[b]{.34\linewidth}
            \includegraphics[width=.9\columnwidth]{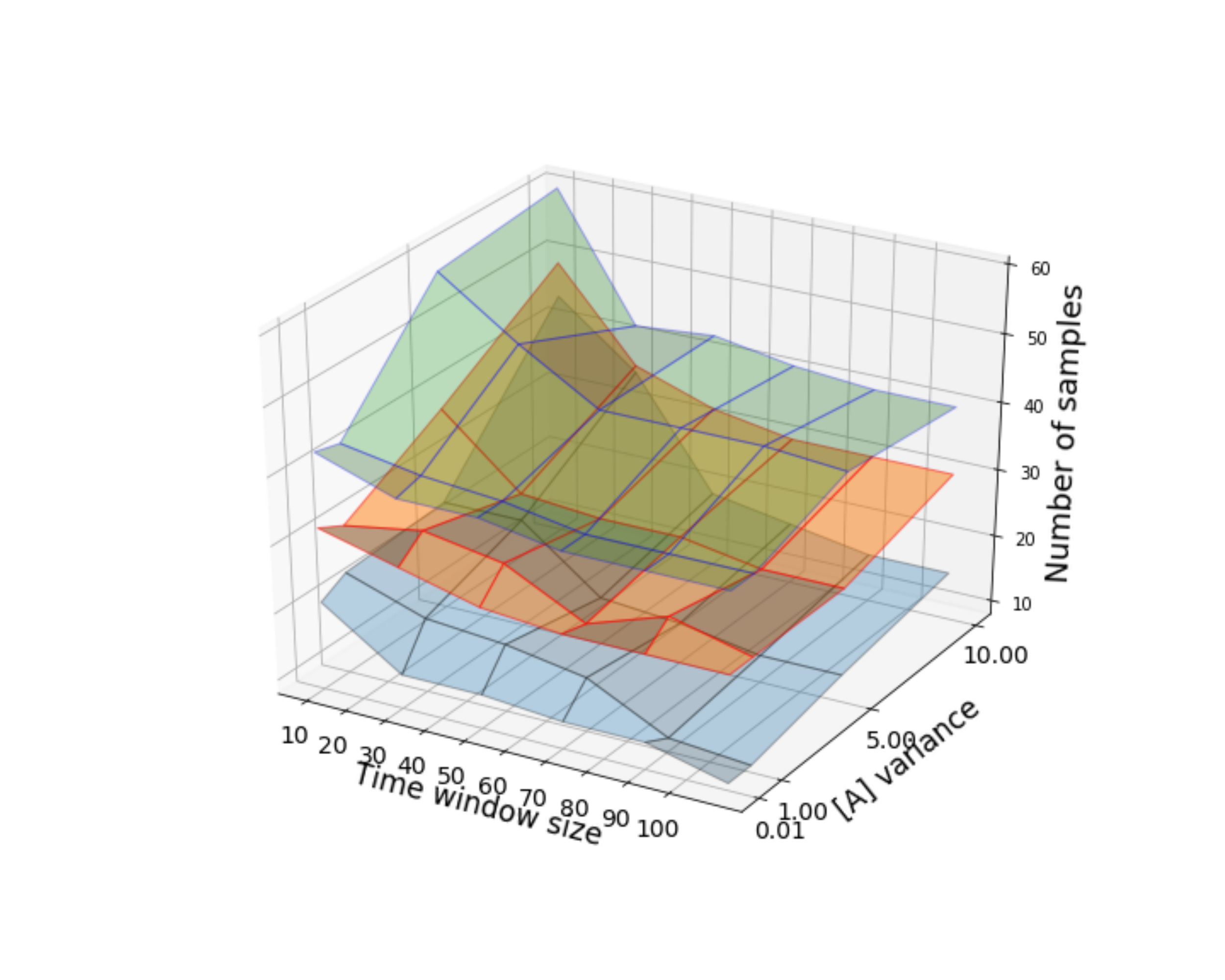}
            \vspace{-1.0em}
            \caption{}
            \label{fig-06-c}
          \end{subfigure}
          \vspace{-1.0em}
          \caption{ Deciding the optimal number of SBU steps ($S_{opt}$) based on, (a)  MSE with variability in $[A]$ and $Tw$, (b) variability in $[A]$ and $Tw$, and (c) MSE with variability in $[A]$; top - $[0,50]$, middle - $[25 - 50]$, and bottom - $[40- 50]$.} 
          \label{fig-06}
        \end{figure*}

\begin{itemize}
    \item [1.] Assuming that the variability in $[A]$ within farms is fixed (say $\sigma= 1$), the inter-farm variability in $[A]$ is fixed in step [a] and varied in step [b], respectively.
    \begin{itemize}
        \item [a.] When $Tw$ is fixed to 50, to illustrate how precisely the system can detect the level of $[A]$, the behavior of MSE in detecting the RC of a set of farms was explored while varying inter-farm $[A]$ over the range $[20, 50]$. Figure \ref{fig-05-a} illustrates that the MSE reduces with the increasing number of samples. Under these settings, at least 12 samples are required ($S_{opt} = 12$) to detect the RC with $p \leq 0.001$ accuracy. 
        
        \item [b.] Figure \ref{fig-05-b} exhibits the $S_{opt}$ required for detecting the RC with increasing $Tw$ size and decreasing inter-farm quantity of $[A]$. It can be seen that $S_{opt}$ is decreasing while increasing $Tw$ and decreasing inter-farm variability range of $[A]$. Therefore, this outcome confirms that when the $Tw$ is large enough and variability in the inter-farm quantity of $[A]$ is less, the level of $[A]$ in a farm can be decided effectively with a fewer samples (or updating steps). This means that the convergence is faster. 
    \end{itemize}

    \item [2.] When $[A]$ is varied within as well as among other farms, Figure \ref{fig-06} exhibits the behavior of the MSE and $S_{opt}$ with intra-farm variability in $[A]$ (denoted as $[A]$ variance) and $Tw$ size. Although the MSE becomes smaller with decreasing variance in $[A]$ regardless of the $Tw$ size (Figure \ref{fig-06-a}), the corresponding $S_{opt}$ is higher for smaller $Tw$s (Figure \ref{fig-06-b}). Furthermore, Figure \ref{fig-06-c} depicts that the optimal $Tw$ size and $S_{opt}$ become higher with larger inter-farm as well as on a specific farm variance in $[A]$. This means that the time taken to detect $[A]$-level with $p \leq 0.001$ accuracy is longer. 
\end{itemize}

\begin{figure}[!t]
  \begin{subfigure}[b]{.48\linewidth}
    \includegraphics[width=.9\columnwidth]{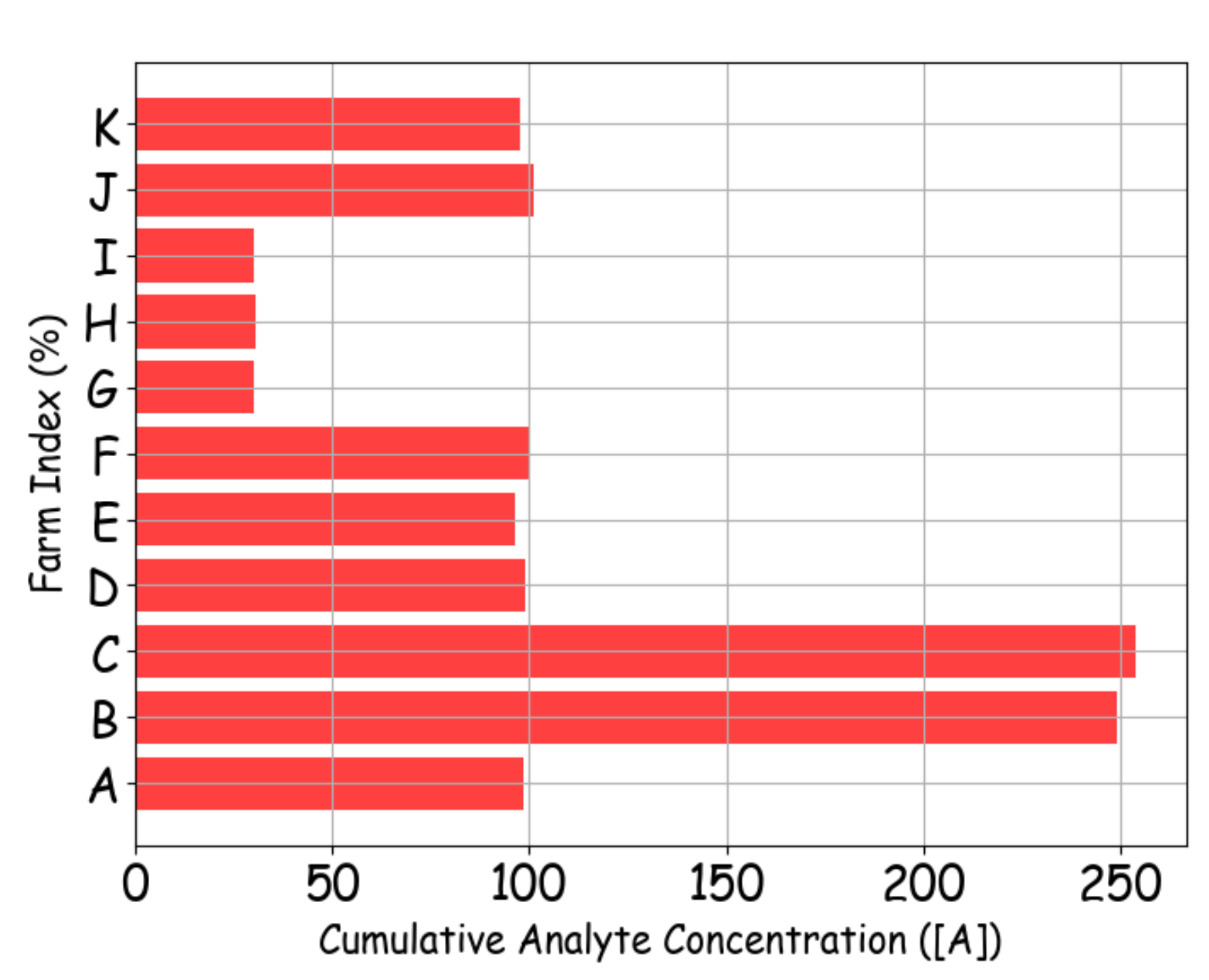}
    \vspace{-0.5em}
    \caption{ }
    \label{fig-07-a}
  \end{subfigure}
  \hfill 
  \begin{subfigure}[b]{.48\linewidth}
    \includegraphics[width=.9\columnwidth]{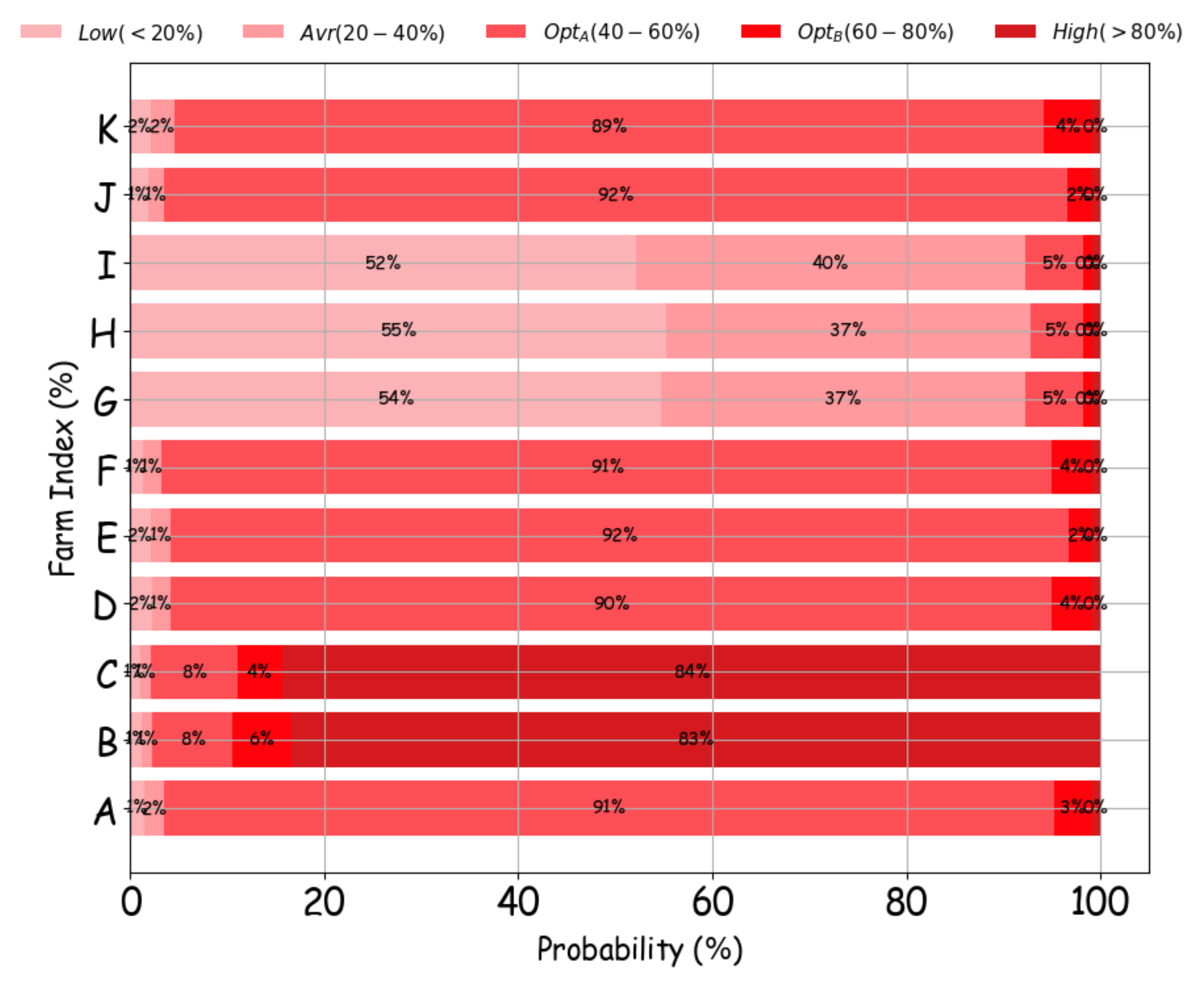}
    \vspace{-0.5em}
    \caption{}
    \label{fig-07-b}
  \end{subfigure}
  \vspace{-0.5em}
  \caption{Color tokens corresponding to the level of [A]; concentration of $A$ (a) and the corresponding color tokens . }
  \label{fig-07}
\end{figure}

Therefore, Figure \ref{fig-05} and \ref{fig-06} reveal that if the system takes longer period at a slow convergence rate for detecting the levels of chemicals, it is an early indication that the variability in the level of that chemical is higher within and/or among farms which are connected through the BC system.

\subsection{Color Tokens and BC System Performance}
We assume in this analysis a chemical $A$ was applied several times at a value of ten over forty farms. The average $[A]$ applied over the farms varied randomly within the range $[0,50]$. Since the average $[A]$ over each farm could vary due to the variability in field conditions such as land usage, it was assumed that the amount of this variability randomly changes over the range $[0,5]$ (i.e., variability in $[A]$ over a farm is the average $[A] \pm p$, where $p\sim\mathcal{U}(0,5)$). Values for $Tw$ and $S_{opt}$ were selected as 50 and 15, respectively, because under similar experimental settings, the previous section showed that with these values, the level of $[A]$ can be detected effectively with $p \leq 0.001$ accuracy.  

At each application of $A$, Algorithm \ref{Al4} was executed to derive the probability of $[A]$ in each farm belonging in the five RCs. The probability values were then converted into a color token. For instance, Figure \ref{fig-07-a} and \ref{fig-07-b} illustrate, respectively, the total amount of chemicals used and the corresponding color token created for ten selected farms. The color token corresponding to the farms B and C, which have used a higher level of chemicals, indicates greater probability ($> 80\%$) of being in the RCs $D$ and $E$. On the other hand, farms which have used least, for instance, G, H, and I, are mostly limited to the RCs $A$ and $B$. Therefore, the color token is a good indicator to represent the levels of the chemicals in the farm.

\begin{figure}[!t]
  \begin{subfigure}[b]{.48\linewidth}
    \includegraphics[width=\columnwidth]{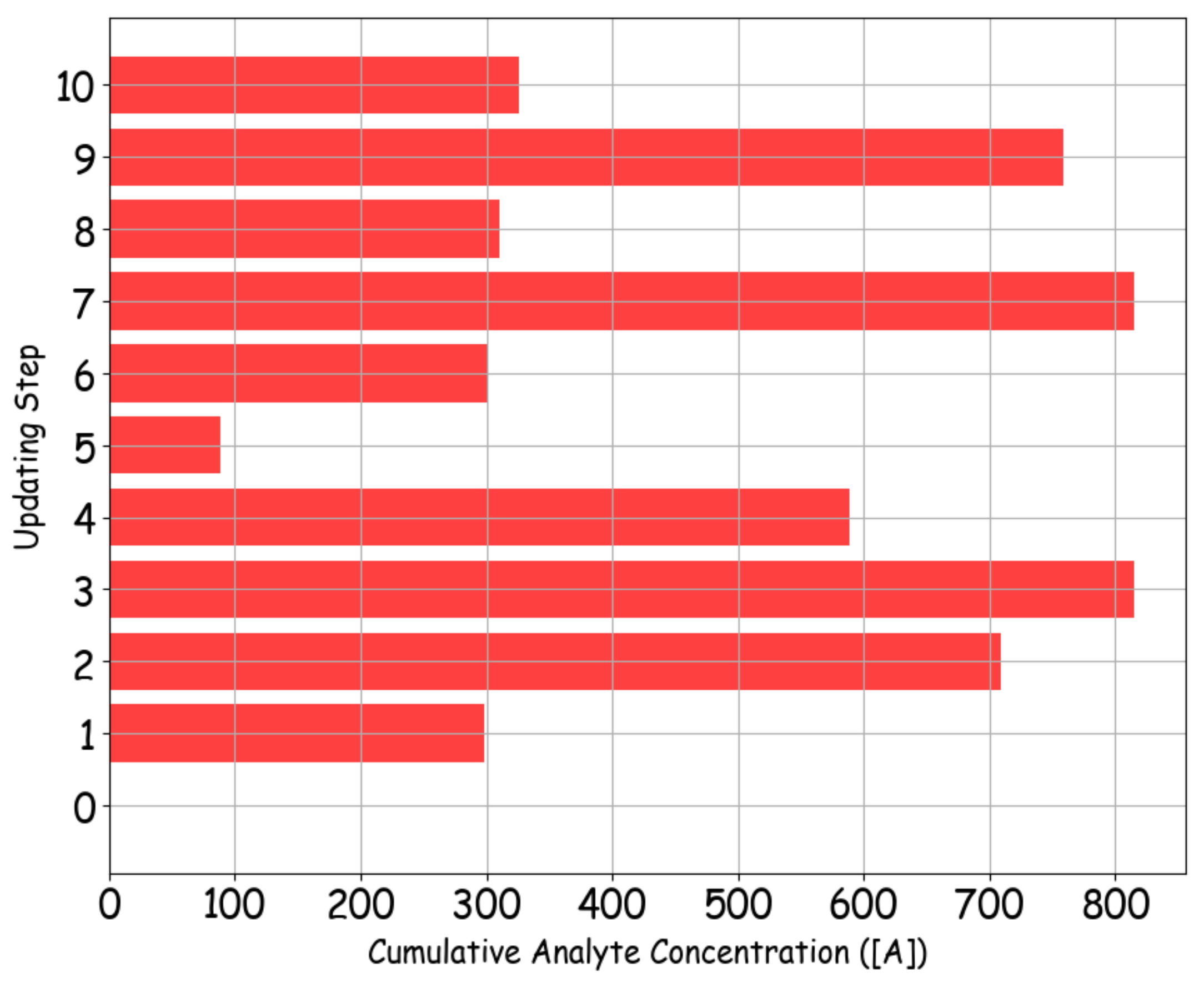}
    \vspace{-1.0em}
    \caption{}
    \label{fig-08-a}
  \end{subfigure}
  \hfill 
  \begin{subfigure}[b]{.48\linewidth}
    \includegraphics[width=\columnwidth]{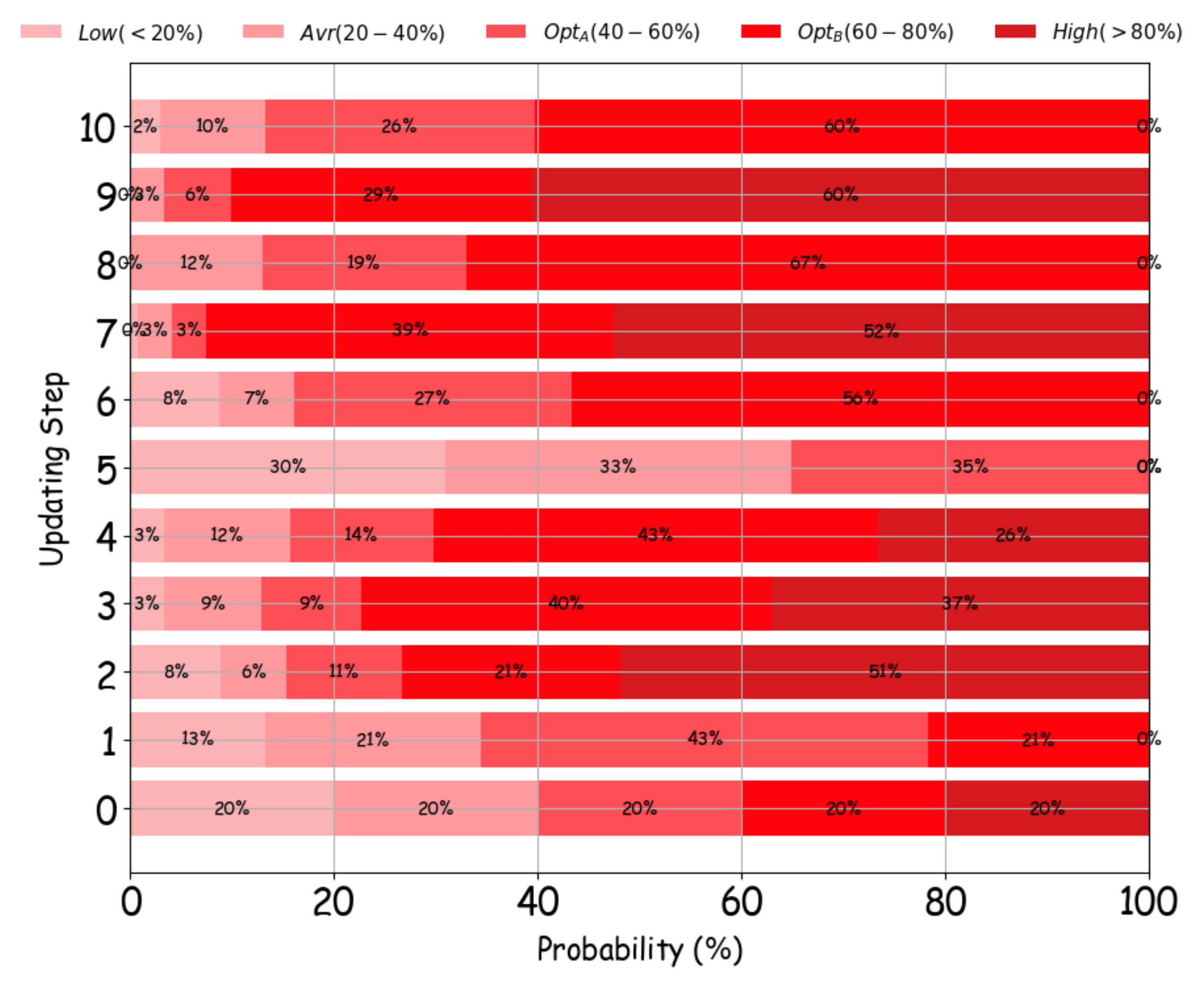}
    \vspace{-1.0em}
    \caption{}
    \label{fig-08-b}
  \end{subfigure}
  \vspace{-1.0em}
  \caption{Color tokens corresponding to overall variability in $[A]$, (a) level of $A$ and (b) color tokens. }
  \label{fig-08}
\end{figure}

Moreover, stakeholders in the supply chain such as policy makers and government bodies may be interested in looking at the overall status of the level of chemicals rather than individual farms. That is because it could enable them to make an overall image of chemical usage and to generate alarming alerts such as over accumulation of certain chemicals which may contribute to increasing GHG emission. It will also help to identify whether there is any impact from external factors such as weather on varying chemical levels in the soil, even though the farmers may claim that a proper amount of chemical has been used. Therefore, to get an overall view on the level of chemicals over the area covered by the BC network, the change in the color tokens over time was explored. Figure\ref{fig-08}, for instance, depicts the color tokens obtained after ten chemical applications, including the corresponding cumulative sum of the chemical used in all forty farms. Similar to Figure \ref{fig-07}, the probability of the overall status of the chemical level being in the RCs $D$ and $E$ is higher with the higher cumulative value of $[A]$. 

\subsubsection{Performance comparison between BC-IoNT and centralized approaches}
\begin{figure}[!t]
        \centering
          \includegraphics[width= .7\linewidth]{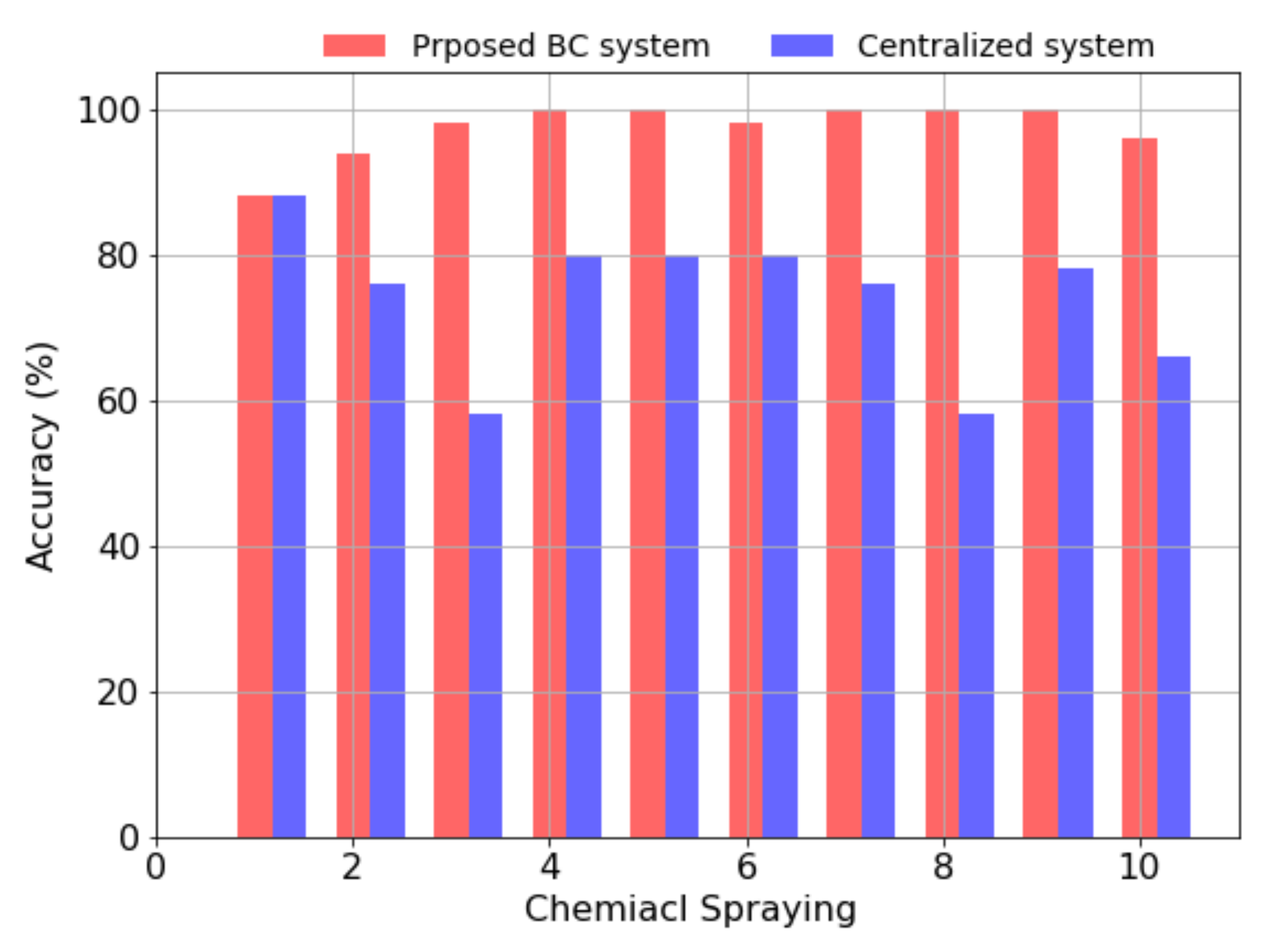}
          \vspace{-1.0em}
          \caption{Accuracy of detecting the chemical levels with the BC-IoNT and centralized approaches for forty farms.}
            \vspace{-1.0em}
        \label{fig-07-c}
        \end{figure}

 The accuracy of detecting the RCs of all forty farms was also computed by using the proposed BC-IoNT system and the centralized approach. Figure \ref{fig-07-c} shows that the proposed system achieves higher accuracy than the centralized approach. That is because the centralized approach independently decides the RC of each farm without taking into account the prior chemical usage at each step. On the other hand, the BC-IoNT uses information in the previous block for creating the current block to represent the chemical concentration status. Therefore, the incorporation of BC into the IoT and IoNT based chemical detection system contributes to improving the accuracy of deciding the chemical levels over a set of farms. 

\section{Discussion} \label{sec5}
In this section, based on the credit transactions between farmers and the miners, traceability of the use of chemicals is first discussed. Second, the credibility of each farm is explored to interpret how well farms are being compliant with chemical standards in the production process, followed by the advantages of the proposed system. 

\subsection{Traceability of the BC system}
The proposed system can be used for ensuring traceability in farm produce. The traceability problems in this case are as (1)traceability in the amount of chemicals detected in a single farm, (2) traceability in the total amount of chemicals used in a farm, and (3) authenticity in the chemicals used in a single farm authentic, and is approved by the regulators.

Our proposed BC-IoNT system can be used to solve these traceability problems as follows: 

\begin{itemize}
    \item [1.] A farmer gets a fixed number of tokens with every purchase of chemicals from a  company. These tokens can be given to the company, along with serial numbers to be used in the serialization of a standard unit of chemical. For example, for $1Kg$ bag of chemicals the company gets an unique serial number and a set of tokens (e.g., say 1000 tokens). After receiving the bags of chemicals and tokens in exchange for another fiat currency \footnote{a kind of national currency that has no intrinsic value and the value depends on the currency issuer such as country's central bank}, leading to the farmer initiating two parallel traceability processes (Figure \ref{fig-09}):
    
        \begin{itemize}
        \item[a.] In the first process, the farmer uses the chemicals as required in their farm. The sensor detects the levels of chemicals and creates a transaction from the farmer's account to the regulator's account to record the levels of chemicals detected.
        
        \begin{figure}[!t]
        \centering
          \includegraphics[width= 1\linewidth]{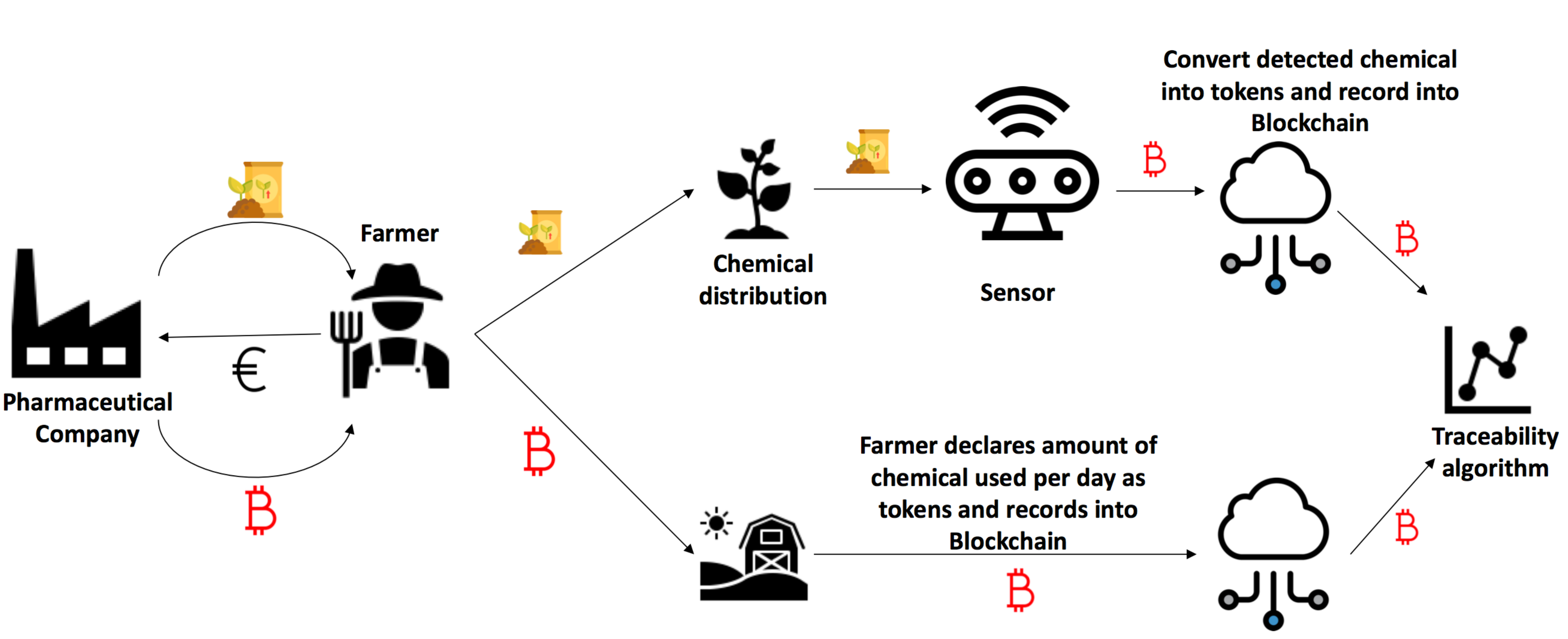}
          \vspace{-2.0em}
          \caption{Traceability assessment process.}
            \vspace{-1.0em}
        \label{fig-09}
        \end{figure}
        
        \item[b.] In the second process, the farmer decides the quantity of chemicals to be use in a day and makes a transaction from their account to a regulator's account with the amount corresponding to amount of chemicals that will be used. For example, if the farmer wants to use 1/10 of a bag of chemicals, this will result in transfer of 100 tokens. 
        \item[c.] A smart contract is executed by the regulators to check if the difference between the amount of tokens in these two processes is within a threshold. The threshold is calculated as per the expected accuracy of sensors.
        \end{itemize}
    \item[2.] This traceability approach can solve the above mentioned traceability problems:
        \begin{itemize}
        \item[1.]    The level of chemicals on a farm can be traced by recording the area of farm from which the item is procured. This procedure will require labeling units of farms, associating each unit of farm with sensors and each farm produce with a set of nearby sensors or the unit of farmland. 
        \item[2.]    If the difference in the amount of tokens in both the traceability process is within a threshold, then it will certify that all chemicals used by the farm is traceable.
        \item[3.]   If the difference in the amount of tokens in both traceability processes is within a threshold, then it will certify that all chemicals used by the farm is authentic. This is because the regulator supplies the tokens.
        \end{itemize}
\end{itemize}

\begin{figure}[!t]
  \begin{subfigure}[b]{.5\linewidth}
    \includegraphics[width=\columnwidth]{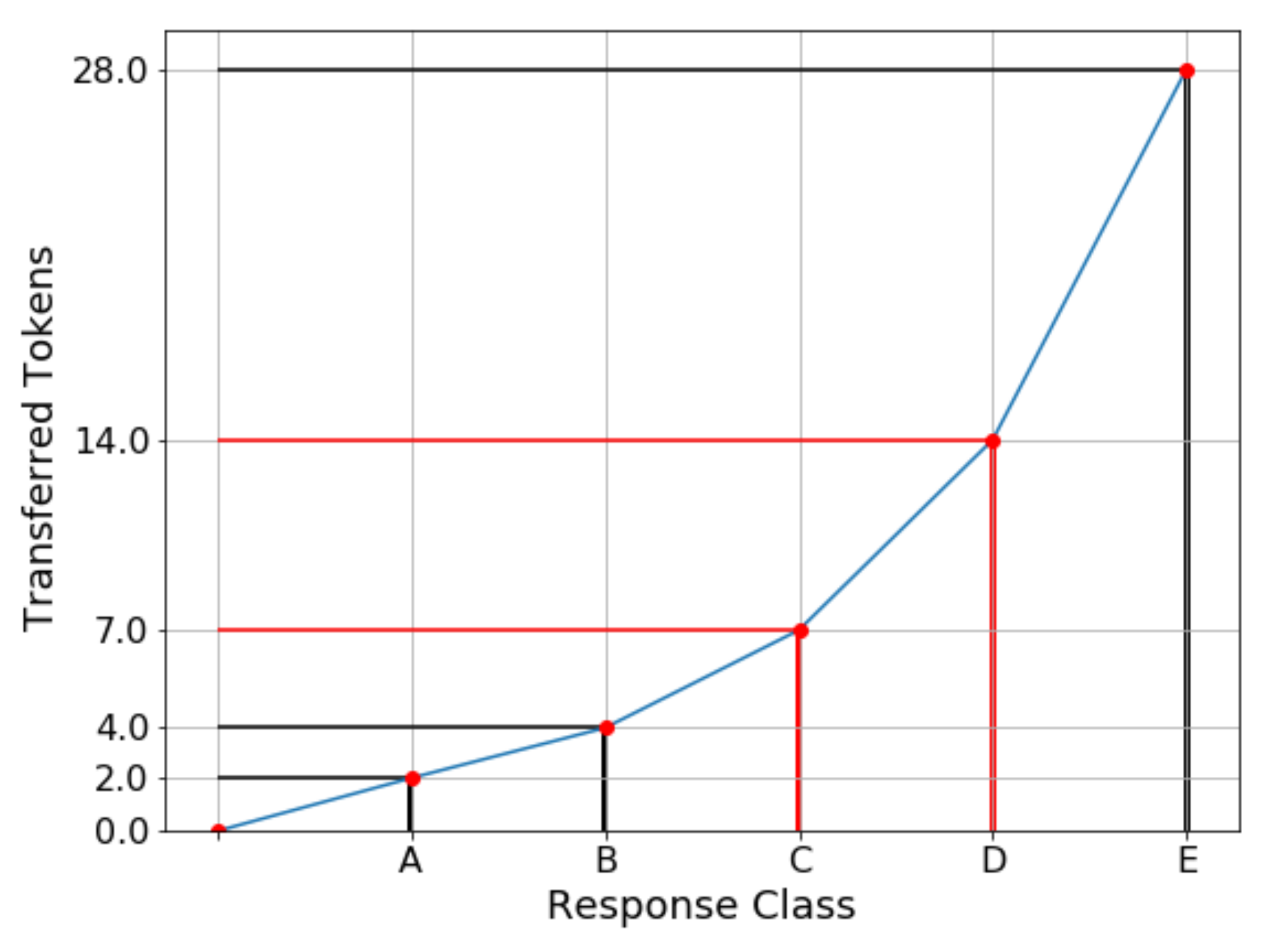}
    \vspace{-1.5em}
    \caption{}
    \label{fig-10-a}
  \end{subfigure}
  \hfill 
  \begin{subfigure}[b]{.5\linewidth}
    \includegraphics[width=\columnwidth]{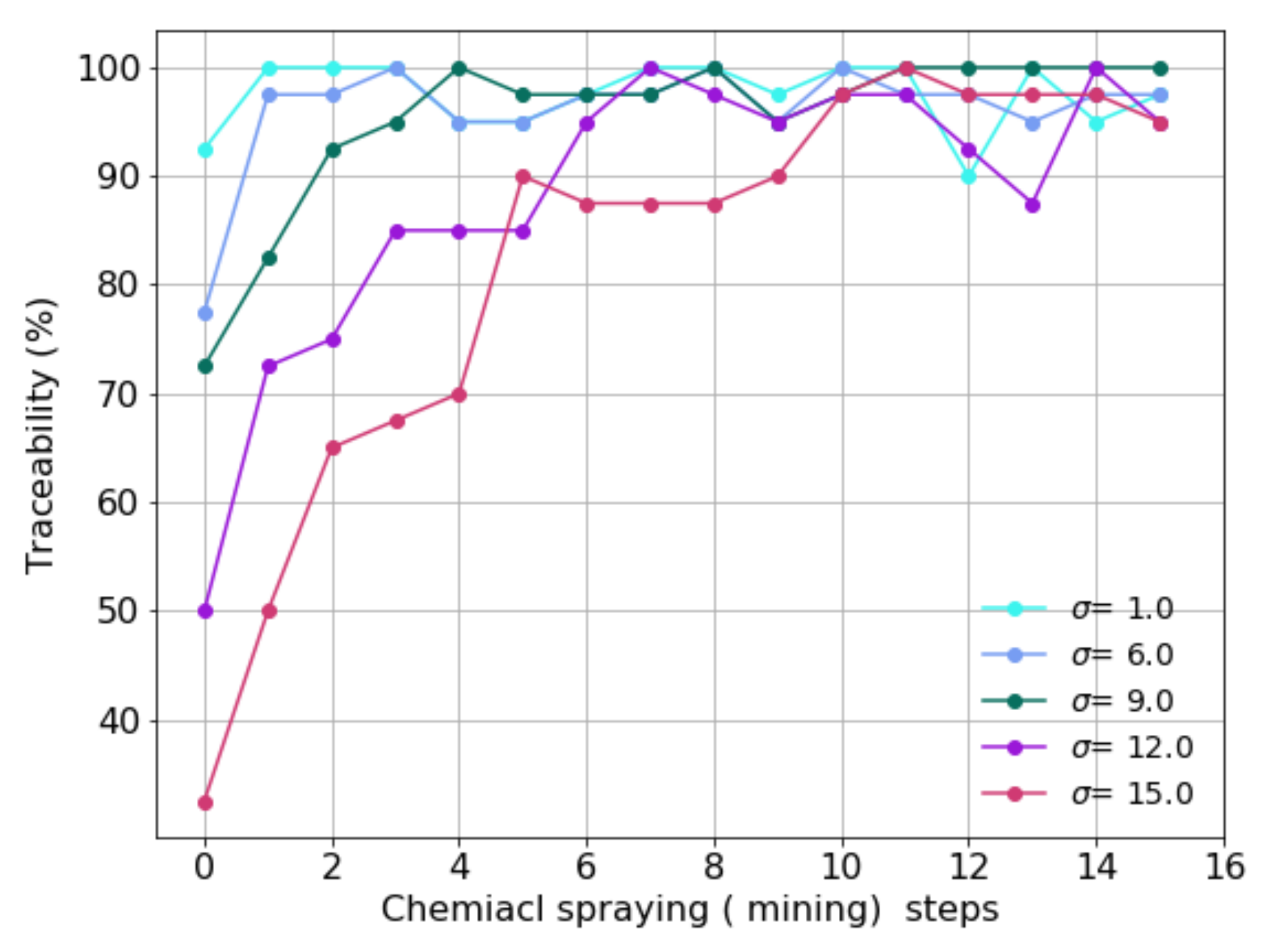}
    \vspace{-1.5em}
    \caption{}
    \label{fig-10-b}
  \end{subfigure}
  \vspace{-1.5em}
  \caption{Traceability evaluation, (a) amount of credits transferred corresponding to RCs and (b) traceability with the variability in $[A]$. }
  \label{fig-10}
\end{figure}

Similar to the credit model explained in section \ref{crdit}, the amount of tokens transferred to the regulator (i.e., mining node) corresponding to the five RCs was exponentially increased as $ e^{\alpha i}$, where $\alpha = 0.05$ and $i = 1, 2, \cdots, 5$ corresponds to the RCs $A, B \cdots, E $. That is because the concentration ranges corresponding to the RCs from $A$ to $E$ increases exponentially (Figure \ref{fig-02-b}). 
By considering Figure \ref{fig-10-a} and \ref{fig-02-b} (i.e., $RF$ with $[A]$) together, the amount of tokens required for different chemical levels can be decided effectively. 

The amount of tokens that each farmer required to send to the regulator was computed based on the RC detected by using his color token. Also, each farmer computed the amounts of tokens based on the level of chemicals used over the farm. The difference between the amount of tokens from these two approaches was computed for forty farms for 15 chemical applications (i.e., mining steps). At each step, the number of farms that achieved the difference in the amount of tokens less than $10^{-3}$ was computed. This token difference could, however, be affected due to the variability in $[A]$ (i.e., $\sigma$) and consequently, traceability can be varied substantially. Therefore, the impact of variability in $[A]$ on the traceability was also explored. Figure \ref{fig-10-b} exhibits the traceability obtained for four different values of $\sigma$. For each $\sigma$, the system achieved at least $\geq 98\%$ traceability though the number of mining steps required for achieving such a higher traceability is increasing with increasing $\sigma$ value.

\subsection{Credibility of farms}
\begin{figure}[!t]
\centering
  \includegraphics[width= .8\linewidth]{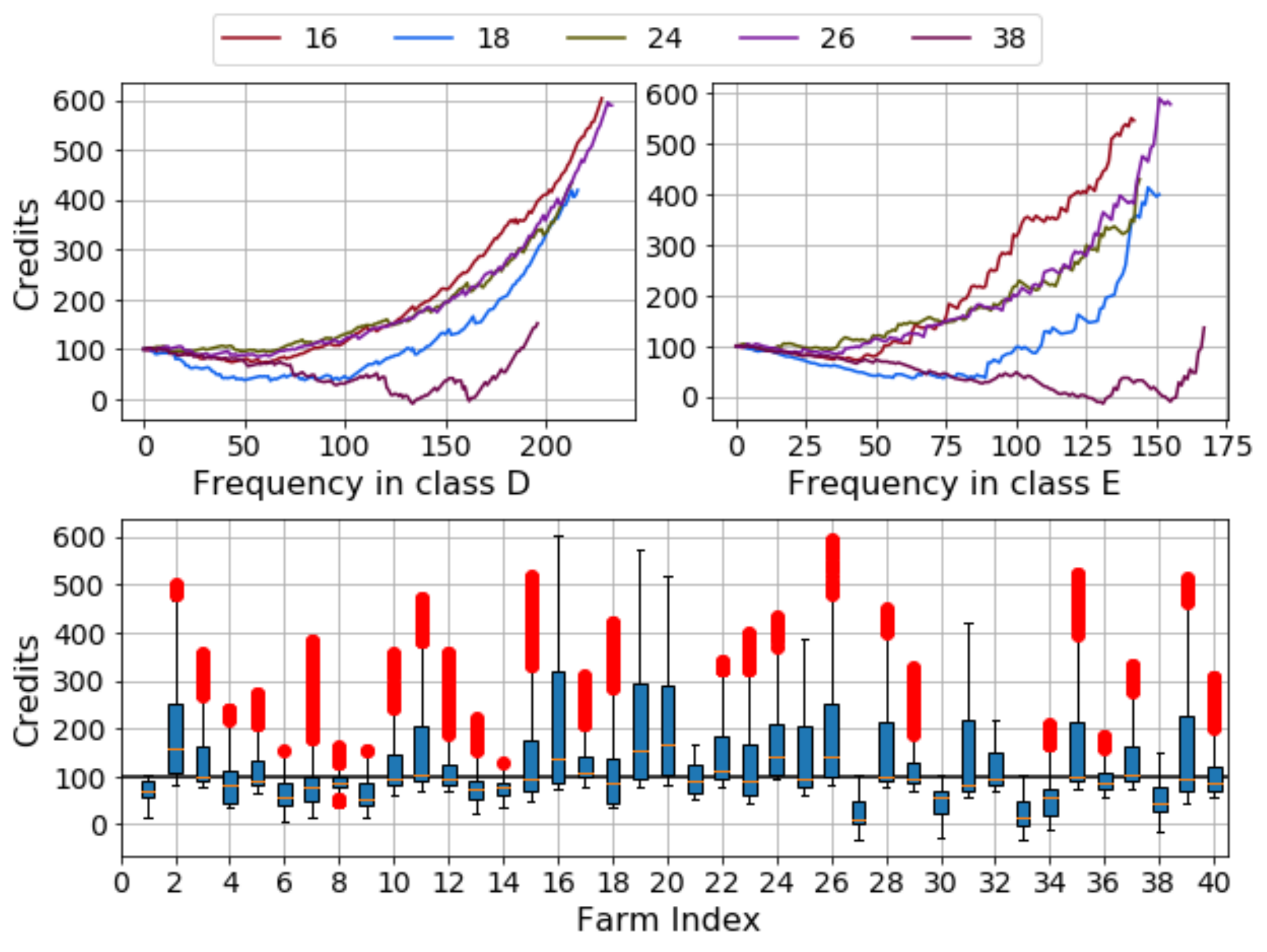}
  \vspace{-1.0em}
  \caption{Variability in credits for selected farms IDs (top) and overall credits (bottom).  }
\label{fig-11}
\end{figure}

 Assuming that initially every farm has 500 credits (i.e., $Cr_0 = 500$) and credits tuning parameter $\alpha = 0.05$, we explored the variability in credibility of each farm by computing the credits of each farm for $T = 1,2,3, \cdots, 100$. Figure \ref{fig-11} depicts the variability in the $Cr$ for selected four farms out of forty farms with their frequency of being in the RCs $D$ and $E$, respectively. The credits earning rate (i.e., steepness of the plots) is high when the frequency of being in the RCs $D$ and $E$ is less. This is because farms earn credits compared to the amount they spend when they are compliant with the chemical standards. Thus, the steepness of the credit curve is an indicator of the credibility. Besides, Figure \ref{fig-11} exhibits the variability in overall credits of forty farms. This helps to obtain a comparable idea about the variability in credibility of a set of farms and also to identify, for instance, the least credible farms easily, and they notify them to manage the use of chemicals.

\begin{figure}[!t]
    \includegraphics[width=.8\columnwidth]{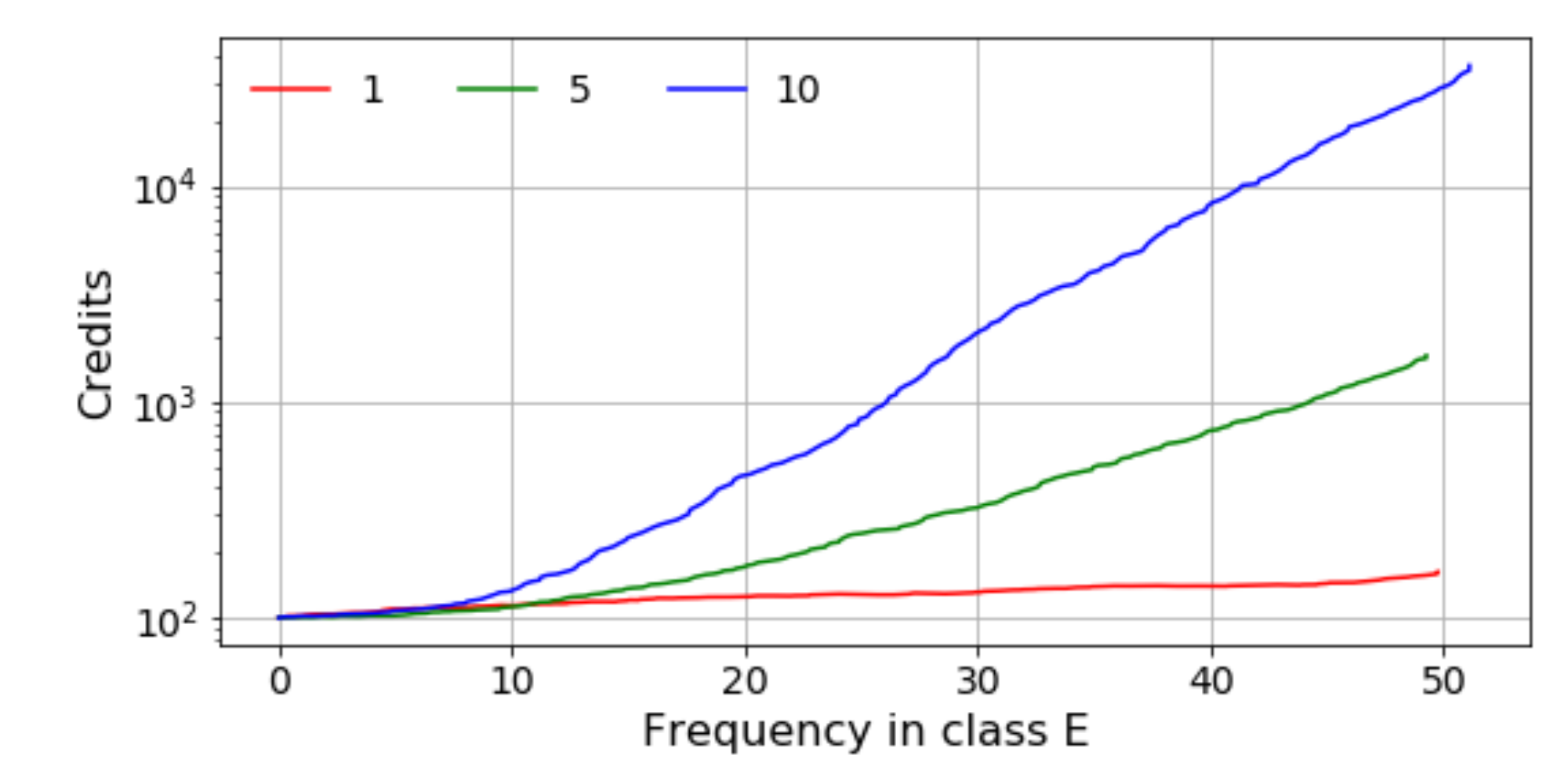}
  \vspace{-1.0em}
  \caption{Change of overall credits with tuning parameter $\alpha$.  }
  \label{fig-12}
\end{figure}

The steepness could, however, be varied with the turning parametr $\alpha$. To understand the impact of $\alpha$, the amount of credits held by the set of forty farms was computed for three different values of $\alpha$. Figure \ref{fig-12} exhibits the behavior of average credits of all forty farms with the average frequency of those farms being in the response class $E$. The steepness of the credit curve increases with increasing $\alpha$. Thus, the credibility of farms can be changed by varying $\alpha$. 

\subsection{Benefits of BC-IoNT}
      This system can be used for effective application of fertilizers. When applying different fertilizers based on their needs, variable-rate fertilization (VF) is one of the recommended and commonly used methods in Smart Farming. The VF method allows optimizing the soil available nutrient levels, ensuring sustainable productivity. It is critically important to have timely accurate information about the soil available nutrient contents in order to use the VF method, though factors such as spatial heterogeneity will limit obtaining this information effectively \cite{9}. In this regard, the VF method combined with the BC-IoNT system will be a promising method to set up an effective fertilization strategy as the BC-IoNT system can be used to detect the soil nutrient content based on molecular-level data. 
    
     The proposed system can also contribute to improving the performance of the food supply chain. Consumers are more conscious about the quality and safety of food due to the lack of transparency in the food supply chain. As a viable solution, the color token generated in the BC-IoNT system can be use used as a reliable indicator (e.g., certificate) to represent the quality of food with respect to the use of chemicals in the food production process (e.g., identifying organic vegetables).  At the same time, authorized parties, such as the FNs in the BC system, can assess the variability in the credibility of farms being compliant with chemical standards and send early warning alerts to producers to maintain an optimal chemical level. This could contribute to reducing farm input, for instance, expenditure on fertilizers and environmental impact. Therefore, the proposed approach has  potential in empowering the validity of such essential features of the food supply.
    
\subsection{Challenges and Future Work}
There are several challenges in incorporating IoNTs into BC-enabled IoT systems that need to be addressed further. Few critical challenges are briefly discussed below, including possible solutions as future works. 

\begin{itemize}
\item {\bf The block propagation speed:} Slow block propagation speed over the BC network creates extended latency in updating the distributed ledgers. This, for example, could be overcome  by integrating the block propagation latency analysis method presented in \cite{3} with the block propagation techniques such as advertisement-based propagation and unsolicited push propagation listed in \cite{26}. 

    \item {\bf Constraints in resources:} Executing computationally heavy smart contracts and space required for storing redundant BC ledger are critical challenges due to limited resources. Hence, employing effective data processing techniques such as compressed sensing \cite{19,20}, integration with Cloud-based BC \cite{4}, and resource shared cooperative computing techniques \cite{23} and mobile-clould offloading \cite{40} would also be promising alternatives. Selecting computing and communication nodes using the importance of nodes in a network explained in \cite{21} would also contribute to utilizing the available resources effectively.

    \item {\bf Effectiveness of smart contracts:} One of the key tasks that can be performed using the proposed BC approach is alarming the highly contaminated farms. The alarming process can be enhanced further by adding dynamics of different factors such as disruption of contaminants in the molecular binding process into Algorithm \ref{Al1} to \ref{Al4}. This would be another extension of this study to improve the timely detection of the level of a chemical. 
    
    \item {\bf Managing security threats:} There are several security threats associated with BC-based systems, and some of the most common threats are private key leakage and vulnerability \cite{25}. In the proposed system, mostly such threats could come from the farmers' side. For example, a farmer may get rid of all blocks and join the BC as a new entity, aiming to avoid bad reputation. Therefore, developing techniques to overcome such issues is crucially important. 
\end{itemize}

\section{Conclusion} \label{sec6}
This study focused on incorporating BC into IoT and IoNT combined system (BC-IoNT) to develop a distributed decision-making system for sensing the level of chemicals used in farms. Then, BC-IoNT was used to facilitate providing more reliable and transparent information that enables making efforts to minimize carbon footprint and also selecting more safe and eco-friendly products for stakeholders in the food production and supply chain. Also, BC-IoNT increases the use of data as stakeholders like farmers can share the chemical usage data more secularly and in turn get benefits such as improved production quality and earn recognition for the products in the market.

The Langmuir molecular binding model and the Bayesian probability updating method based ML model was used as a smart contract in the system, which can effectively sense the levels of chemicals. This information is then shared as a color token over the BC network.  The data analysis confirmed that BC-IoNT system detected the level of chemicals in farms with higher accuracy than the centralized approach. The study concluded further that the efficiency of detecting the levels of chemicals could be varied due to several parameters such as variability in chemical concentration within as well as among farms. Based on this, our study found that selecting the optimal sampling frequency and the optimal number of probability updating steps is critical to improve the efficiency of the system in detection of the level of chemicals in farms. Moreover, the rate of change in the amount of credits held by each farm can be used as an indicator of the credibility to reward farms that are compliant with the recommended chemical standards (i.e., being eco-friendly). Similarly, the token-based traceability method confirmed that the BC-based system could achieve higher traceability, but the time taken for that could be varied with the variability in the level of chemicals over the farms. 
 
This study is, however, a new research direction as the integration of IoNT and IoT into BC-enabled decision-making systems have not been considered before. Therefore, this system can bring several advantages to several application domains and also lays the foundation for several future research directions as there are many challenges that need further attention.

\section*{Acknowledgment}

This research was supported by the Science Foundation Ireland (SFI) project ”PrecisionDairy (ID: 13/1A/1977)” as well as a research grant from Science Foundation Ireland and the Department of Agriculture, Food and Marine on behalf of the Government of Ireland under the Grant 16/RC/3835 (VistaMilk). 

\bibliography{MainScript}

\bibliographystyle{IEEEtran}

\end{document}